\documentclass[12pt]{article}

\usepackage[fleqn]{amsmath}
\usepackage{amsfonts}
\usepackage{amssymb}
\usepackage{color}
\usepackage{epsfig}
\usepackage{geometry}

 \usepackage[super,compress]{cite}
\usepackage{cite}

\usepackage{graphicx}
\usepackage{float}

\newtheorem{thm}{THEOREM}[section]

\newtheorem{rem}[thm]{Remark}

\newcommand{\Rset}{\mathbb{R}}
\newcommand{\sV}{{\vect{s}}}            
            
\newcommand{\nullV}{\pmb{{\cal O}}}
\newcommand{\BV}{\pmb{{\cal B}}}

\newcommand{\DV}{\pmb{{\cal D}}}
\newcommand{\EV}{\pmb{{\cal E}}}
\newcommand{\nab}{\vect{\nabla}}
\newcommand{\alphaS}{\alpha_{\mbox{\tiny{S}}}}
\newcommand{\HV}{\pmb{{\cal H}}}

\newcommand{\pddt}{\frac{\partial\phantom{t}}{\partial t}}
\newcommand{\qV}{{\vect{q}}}

\newcommand{\veps}{\varepsilon}
\newcommand{\drm}{\mathrm{d}}
\newcommand{\refeq}[1]{(\ref{#1})}
\newcommand{\vect}[1] {\boldsymbol{{ #1}} }
\newcommand{\Nset}{\mathbb{N}}
\newcommand{\Sset}{\mathbb{S}}
\newcommand{\QV}{{\vect{Q}}}            
\newcommand{\AV}{\pmb{{\cal A}}}
\newcommand{\cH}{{\mathcal{H}}}
\newcommand{\lambdaC}{\lambda_{\mbox{\tiny{C}}}}

\newcommand{\mEL}{{m}_{\rm{el}}}       
\newcommand{\mPR}{{m}_{\rm{pr}}}       

\newcommand{\rPR}{{r}_{\rm{pr}}}      

\begin{document}

\markboth{Holly Carley, Michael Kiessling, Volker Perlick}
{Schr{\"o}dinger spectrum of a hydrogen atom with
Bopp-Land{\'e}-Thomas-Podolsky interaction}

\title{ON THE SCHR{\"O}DINGER SPECTRUM OF A HYDROGEN ATOM WITH ELECTROSTATIC BOPP--LAND{\'E}--THOMAS--PODOLSKY INTERACTION BETWEEN ELECTRON AND PROTON}

\title{\uppercase{On the Schr\"odinger spectrum of \\ a hydrogen atom with \\ electrostatic \\
\uppercase{B}{\small\sc{opp}}--\uppercase{L}{\small\sc{and\'e}}--\uppercase{T}{\small\sc{homas}}--\uppercase{P}{\small\sc{odolsky}}\\
interaction\\ between electron and proton}}
\author{\textbf{Holly K. Carley,$^1$ Michael K.-H. Kiessling,$^2$ Volker Perlick$^3$}\\
                $^1$ Department of Mathematics, \\
                New York City College of Technology, CUNY\\
                 300 Jay Street, Brooklyn, New York, NY 11201, USA\\
                 $^2$ Department of Mathematics, Rutgers University\\
                110 Frelinghuysen Rd., Piscataway, NJ 08854, USA\\ 
                $^3$ ZARM, Universit\"at Bremen, 28359 Bremen, Germany \\
\textrm{\small Version of Sept. 03, 2019. Typeset with \LaTeX\ on: }}
\maketitle

\vspace{4truecm}
\copyright(2019) \small{The authors. Reproduction of this preprint, in its entirety, is 

permitted for non-commercial purposes only.}

\thispagestyle{empty}

\begin{abstract}
The Schr\"odinger spectrum of a hydrogen atom, modelled as a two-body system consisting 
of a point electron and a point proton, changes when the usual Coulomb interaction between 
point particles is replaced with an interaction which results from a modification of 
Maxwell's law of the electromagnetic vacuum. Empirical spectral data thereby impose 
bounds on the theoretical parameters involved in such modified vacuum laws. In the 
present paper the vacuum law proposed, in the 1940s, by Bopp, Land\'e--Thomas, and 
Podolsky (BLTP) is scrutinized in such a manner. The BLTP theory hypothesizes the 
existence of an electromagnetic length scale of nature --- the Bopp length 
$\varkappa ^{-1}$ ---, to the effect that the electrostatic pair interaction deviates 
significantly from Coulomb's law only for distances much shorter than $\varkappa^{-1}$.
Rigorous lower and upper bounds are constructed for the Schr\"odinger energy levels of 
the hydrogen atom, $E_{\ell, n}(\varkappa)$, for all $\ell\in\{0,1,2,...\}$ and $n >\ell$.
The energy levels $E_{0,1}(\varkappa)$, $E_{0,2}(\varkappa)$, and $E_{1,2}(\varkappa)$ 
are also computed numerically and plotted versus $\varkappa ^{-1}$. It is found that 
the BLTP theory predicts a non-relativistic correction to the splitting of the 
Lyman-$\alpha$ line in addition to its well-known relativistic fine-structure splitting. 
Under the assumption that this splitting doesn't go away in a relativistic calculation, 
it is argued that present-day precision measurements of the Lyman-$\alpha$ line suggest 
that $\varkappa ^{-1}$ must be smaller than $\approx 10^{-18} \, \mathrm{m}$. 
Finite proton size effects are found not to modify this conclusion. As a 
consequence, the electrostatic field energy of an elementary point charge, although 
finite in BLTP electrodynamics, is {much larger than the empirical rest mass 
($\times \, c^2$) of an electron}. If, as assumed in all ``renormalized theories'' of 
the electron, the empirical rest mass of a physical electron is the sum of its bare rest 
mass and its electrostatic field energy, then in BLTP electrodynamics the electron has 
to be assigned a negative bare rest mass.
\end{abstract}

\maketitle
\newpage	
\section{Introduction}

Maxwell's ``law of the pure ether,'' 
\begin{eqnarray}
        \HV(t,\sV)  
&= \label{eq:MlawBisHintro}
        \BV(t,\sV) ,
\\
        \DV(t,\sV)  
&=
        \EV(t,\sV) ,
\label{eq:MlawEisDintro}
\end{eqnarray}
has long been suspected as culprit for the notorious electromagnetic ``self''-interaction 
problems of point charges. Of course, the issue is not the long obsolete original notion 
of electromagnetic fields as mathematical expressions of ``stresses in an elastic ether'' 
--- this is simply taken care of by thinking of the postulated identities 
(\ref{eq:MlawBisHintro}), (\ref{eq:MlawEisDintro}) as ``Maxwell's law of the 
electromagnetic vacuum.'' The issue is that this electromagnetic vacuum law leads to 
the Maxwell--Lorentz equations for the electromagnetic fields with point charge sources, 
the solutions of which diverge so strongly at the locations of the point charges that
their electromagnetic field energy, field momentum, and field angular momentum are all 
infinite, and so is the total Lorentz force. Since the first-order PDEs of Maxwell's 
theory linking the fields $\BV,\DV,\EV,\HV$ are a mathematical consequence of the law 
of charge conservation, these PDEs are unlikely to be at fault, which thus points to
the law of the electromagnetic vacuum as the prime suspect.

In the 1940s, Bopp\cite{Bopp}, independently Land\'e and Thomas\cite{landethomas}, and 
subsequently Podolsky\cite{Podolsky} (BLTP) proposed the electromagnetic vacuum law 
\begin{eqnarray}
        \HV(t,\sV)  
&= \label{eq:BLTPlawBandHintro}
       \left(1  + \varkappa^{-2}\square\,\right) \BV(t,\sV) \, ,
\\
        \DV(t,\sV) 
&=
        \left(1  + \varkappa^{-2}\square\,\right) \EV(t,\sV)\, ,
\label{eq:BLTPlawEandDintro}
\end{eqnarray}
to cure the above-mentioned infinities. Here, $\square \equiv c^{-2}\partial_t^2 -\Delta$ 
is the d'Alembertian, and $1/\varkappa$ is ``Bopp's length parameter''\cite{Bopp}; then 
$\varkappa\to\infty$ the BLTP law \refeq{eq:BLTPlawBandHintro}, 
\refeq{eq:BLTPlawEandDintro} reduces to Maxwell's law (\ref{eq:MlawBisHintro}), 
(\ref{eq:MlawEisDintro}). Whether the BLTP vacuum law (\ref{eq:BLTPlawBandHintro}), 
(\ref{eq:BLTPlawEandDintro}) leads to an acceptable classical electrodynamics is 
an interesting question which has also attracted the attention of some of the present 
authors, see Refs. \citen{GratusETal} and \citen{mikishadi}.

In this paper we are concerned with some of its quantum-physical implications. 
 In 1960 Kvasnica \cite{Kvasnica1960} estimated how small $\varkappa$ would have 
to be for explaining an apparent discrepancy between the observed and the predicted Lamb 
shift. Cuzinatto et al. \cite{Cuzinattoetal2011} used the Rayleigh-Ritz method to estimate
the ground-state energy of hydrogen in dependence of $\varkappa^{-1}$. 
 A more systematic study of the complete hydrogen spectrum, assuming a BLTP vacuum law, seems not
yet to have been done.

In the ensuing sections we will investigate how the BLTP vacuum law 
(\ref{eq:BLTPlawBandHintro}), (\ref{eq:BLTPlawEandDintro}) affects the Schr\"odinger 
spectrum of hydrogen. This question can be studied in great detail, for the electric 
BLTP pair interaction energy $V_\varkappa(r)$ of a point electron and a point proton 
a distance $r$ apart is explicitly computable as $V_\varkappa(r) = - e^2 
\frac{1-e^{-\varkappa r}}{r}$; when $\varkappa\to\infty$ this BLTP pair interaction 
energy reduces to the usual Coulomb pair energy $V_\infty(r) = - e^2 \frac{1}{r}$. 
Thus we expect a small perturbation of the usual Rydberg spectrum for large $\varkappa$, 
but significant deviations for when $\varkappa$ becomes too small. 

In the next section we formulate the Schr\"odinger two-body problem with BLTP pair 
interaction energy, reduce it to a single ODE problem in the radial variable for the 
eigenvalues of the relative Hamiltonian, then obtain rigorous upper and lower 
bounds on the eigenvalues, and finally compute the few lowest eigenvalues 
numerically.\footnote{We remark in passing that the Schr\"odinger potential 
   $V_\varkappa(r)$ is a special case of the Hellmann potential $ - \, 
   \frac{A - B e^{- \varkappa r}}{r}$, see Ref. \citen{Hellmann1935}, cf. Ref.   
   \citen{Hellmann1937}, here with $A= B = e^2$. The Hellmann potential
   reduces to the Coulomb potential for $B=0$ and to the Yukawa potential for $A=0$. It 
   is often used in chemistry. Various approximation methods have been worked
   out for determining the energy levels. If $B \ll A$, standard perturbation theory 
   can be applied straightforwardly, but not so in the case $A=B$ which is of interest to us. 
   Methods for numerically calculating the energy levels of the Hellmann potential 
   have been worked out by Adamowski \cite{Adamowski1985} (cf. Amore and 
   Fern{\'a}ndez \cite{AmoreFernandez2014}) and by Vanden Berghe et 
   al. \cite{VandenBergheFackMeyer1989}.\label{fn:ONE}}
Comparison with experimental spectral results\footnote{For experimental bounds on 
$\varkappa$ obtained with other (i.e., non-spectroscopic) methods we refer to Bonin et 
al. \cite{BoninEtAl2010} and to Accioly and Scatena \cite{AcciolyScatena2010}.} 
yields an empirical upper bound on $\varkappa^{-1}$ of approximately $10^{-18} \, \mathrm{m}$, 
see section \ref{sec:Viability}. This is much smaller than the empirical 
proton radius, which roughly coincides with the so-called ``classical electron radius''
$e^2/\mEL c^2$. 
Section \ref{sec:SummaryOutlook} concludes with a summary, and some open questions.

\section{Nonrelativistic ``BLTP hydrogen''} \label{sec:BLTPhydrogen} 

\subsection{Maxwell--Bopp--Land\'e--Thomas--Podolsky field theory}\label{sec:MBLTPtheory}

For the full \emph{Maxwell-BLTP field theory} we refer the reader to the original articles 
Refs. \citen{Bopp}, \citen{landethomas}, and \citen{Podolsky}, as well as 
Ref. \citen{PodolskySchwed}, and recently Refs. \citen{GratusETal} 
and \citen{mikishadi}. Here we recall the Maxwell-BLTP 
field equations specialized to the needs of our investigation, i.e. the (electro-)static 
fields of an arbitrary static configuration of two elementary point charges, one of them 
positive (representing the hydrogen nucleus), the other one negative (representing the 
electron). We then state and evaluate the field energy of such an arbitrary static 
configuration, obtaining a sum of finite self-field energy terms plus the pair interaction 
energy term.

\subsubsection{The static field equations}\label{sec:MBLTPeqns}

The differential equations of the  relativistic field theory will be written 
w.r.t. any particular flat foliation of Minkowski spacetime into space points 
$\sV\in\Rset^3$ at time $t\in\Rset$. In the static limit, $\pddt\BV =\nullV = 
	\pddt\DV$, as well as $\dot{\qV}{}_\pm =\nullV$; here, $\qV_\pm^{}\in\Rset^3$ are the positions of 
proton ($_+$) and electron ($_-$), and  $\dot{\qV}{}_\pm$ their velocities.
 Therefore Maxwell's two evolution equations for the magnetic induction field $\BV$ and the electric displacement field $\DV$ 
reduce to
\begin{eqnarray}
\textstyle
\nab\times\EV(\sV) 
&= \label{eq:MdotB}
\nullV \, ,
\\
\textstyle
\nab\times\HV(\sV)  
&= 
\nullV \, ,
\label{eq:MdotD}
\end{eqnarray}
while Maxwell's constraint equations for these two fields, given two point charges, read
\begin{eqnarray}
&        \nab\cdot \BV(\sV)  
 = \label{eq:MdivB}
       0\, ,
\\
&       \nab\cdot\DV(\sV)  
 =
       4 \pi e \left( \delta_{\qV_+^{}}(\sV)-\delta_{\qV_-^{}}(\sV)\right)\, .
\label{eq:MdivD}
\end{eqnarray}
 The two equations of the ``BLTP law of the electromagnetic vacuum'' reduce to 
\begin{eqnarray}
        \HV(\sV)  
&= \label{eq:BLTPlawBandH}
       \left(1  - \varkappa^{-2}\Delta\,\right) \BV(\sV) \, ,
\\
        \DV(\sV) 
&=
        \left(1  - \varkappa^{-2}\Delta\,\right) \EV(\sV) \, ;
\label{eq:BLTPlawEandD}
\end{eqnarray}
here, $\Delta$ is the Laplacian. 

\subsubsection{The static field solutions}\label{sec:MBLTPeqnsSOLVED}

The static Maxwell-BLTP field equations are easily solved uniquely at all space points 
$\sV\in\Rset^3$ for the asymptotic conditions that $\BV(\sV)$, $\EV(\sV)$, 
$\DV(\sV)$, $\HV(\sV)$ vanish as $|\sV|\to\infty$.  Namely, because of (\ref{eq:MdivB}) 
we can set $\BV(\sV) =\nabla\times \AV(\sV)$ with $\nabla\cdot\AV(\sV)=0$ and 
$\AV(\sV)\to0$ as $|\sV|\to\infty$, while (\ref{eq:MdotB}) implies that we 
can set $\EV(\sV)=-\nabla\phi(\sV)$ with $\phi(\sV)\to0$ as $|\sV|\to\infty$.
Inserting these representations of $\BV$ and $\EV$ into the r.h.s.s of 
(\ref{eq:BLTPlawEandD}) and (\ref{eq:BLTPlawBandH}), then hitting  
(\ref{eq:BLTPlawEandD}) with $\nabla\cdot$ and (\ref{eq:BLTPlawBandH}) with 
$\nabla\times$, and finally using 
(\ref{eq:MdotD}) and (\ref{eq:MdivD}), we obtain
\begin{eqnarray}
&      - \left(1  - \varkappa^{-2}\Delta\,\right) \Delta\AV(\sV) 
= \label{eq:BLTPlawA}
        \nullV
\, ,
\\
&      - \left(1  - \varkappa^{-2}\Delta\,\right) \Delta\phi(\sV) 
=
        4 \pi e \left( \delta_{\qV_+^{}}(\sV)-\delta_{\qV_-^{}}(\sV)\right) , \: 
\label{eq:BLTPlawPHI}
\end{eqnarray}
together with the asymptotic conditions that $(\phi,\AV)$ as well as 
$\Delta(\phi,\AV)$ go to 0 as $|\sV|\to\infty$. The unique solutions are 
\begin{eqnarray}
 \phi(\sV) 
=
 e \left( \frac{1-e^{-\varkappa |\sV - \qV_+^{}|}}{|\sV - \qV_+^{}|}
-
\frac{1-e^{-\varkappa |\sV - \qV_-^{}|}}{|\sV - \qV_-^{}|}\right),
\label{eq:BLTPphiTWOpt}
\end{eqnarray}
and $\AV(\sV)\equiv \nullV$, yielding $\BV(\sV)\equiv \nullV\equiv \HV(\sV)$, 
which is all we need to compute the pair interaction energy.

\subsubsection{The electrostatic field energy}\label{sec:MBLTPfieldENERGY}

In electrostatic situations, the Maxwell-BLTP \emph{field energy density}  
$\veps_{\mbox{\tiny{field}}}(\sV)$ is given by 
\begin{eqnarray}
 \veps_{\mbox{\tiny{field}}} (\sV)
= \label{eq:TooMBLTP}
\frac{1}{4\pi}\left(\EV\cdot\DV  - \frac{1}{2} |\EV|^2 -  
\frac{1}{2\varkappa^2}  \big(\nabla\cdot\EV\big)^2 \right)(\sV). \quad
\end{eqnarray}
Integrating \refeq{eq:TooMBLTP} over $\Rset^3$ yields the electrostatic field energy\footnote{An integration 
by parts, and a rewriting with the help of the Maxwell-BLTP field equations, yields the familiar expression
$\int_{\Rset^3} \veps_{\mbox{\tiny{field}}} (\sV) d^3s = 
\frac{1}{8\pi}\int_{\Rset^3}{\EV\cdot\DV} d^3s$ for the electrostatic field energy.
Similarly one can show that the field energy of a \emph{static} electromagnetic 
Maxwell-BLTP field is given by $\int_{\Rset^3} \veps_{\mbox{\tiny{field}}} (\sV) d^3s = 
\frac{1}{8\pi}\int_{\Rset^3}\left({\EV\cdot\DV}+ {\BV\cdot\HV}\right) d^3s$. This 
familiar identity does not hold for \emph{dynamical} Maxwell-BLTP fields.} $\int_{\Rset^3} 
\veps_{\mbox{\tiny{field}}} (\sV) d^3s =: E_{\mbox{\tiny{field}}}(\qV_+^{},\qV_-^{})$ 
of the two point charges as 
\begin{eqnarray}
E_{\mbox{\tiny{field}}}(\qV_+^{},\qV_-^{}) 
= \label{eq:TooMBLTPint}
e^2 \varkappa - e^2\; \frac{1-e^{-\varkappa |\qV_+^{} - \qV_-^{}|}}{|\qV_+^{} 
- \qV_-^{}|}.
\end{eqnarray}
Here, the configuration-independent term is just the sum of the electrostatic 
self-field energies of the two point charges of magnitude $e$, which amounts to 
twice the self-field energy of a single point charge of magnitude $e$, and which 
is manifestly finite in the BLTP theory.
 The configuration-dependent term is the 
pair-interaction energy between a positive and a negative point charge, each of 
magnitude $e$.

\subsection{\hspace{-5pt}The Schr\"odinger Equation for ``BLTP Hydrogen''} \label{sec:BLTPhydrogenERWIN}

Ignoring the constant electrostatic self-field energy terms, the Schr\"odinger 
equation for the joint wave function $\Psi(t,\qV_+^{},\qV_-^{})$ of the electron and 
the proton in a ``BLTP hydrogen'' atom reads
\begin{eqnarray}
 i\hbar\pddt\Psi = \label{eq:ErwinEQ} H \Psi
\end{eqnarray}
with
\begin{eqnarray}\label{eq:H}\hspace{-10pt}
H := -\frac{\hbar^2}{2m_+^{}}\Delta_+^{}
-\frac{\hbar^2}{2m_-^{}}\Delta_-^{}
 - e^2 \frac{1-e^{-\varkappa |\qV_+^{} - \qV_-^{}|}}{|\qV_+^{} - \qV_-^{}|};\quad
\end{eqnarray}
here, $m_+^{}=\mPR$ is the proton's and $m_-^{} = \mEL$ the electron's rest mass, and 
$\hbar$ is Planck's quantum of action divided by $2\pi$. This equation can be treated 
with the same separation-of-variables technique which solves the traditional textbook 
problem of the Schr\"odinger spectrum for hydrogen, as follows.

\subsubsection{Separating Center-of-Mass from Relative Coordinates} \label{sec:CMandREL}

We define the center-of-mass coordinate
\begin{eqnarray}
\QV
= \label{eq:CM}
\frac{m_+^{}\qV_+^{} + m_-^{}\qV_-^{}}{m_+^{}+m_-^{}}
\end{eqnarray}
and the relative position vector
\begin{eqnarray}
\qV
= \label{eq:REL}
\qV_-^{} -\qV_+^{}.
\end{eqnarray}
Setting $m_+^{}+m_-^{} =M$ and $\frac{m_+^{}m_-^{}}{m_+^{}+m_-^{}}=\mu$, 
(\ref{eq:ErwinEQ}) becomes
\begin{eqnarray}
\hspace{-10pt}
i\hbar\pddt\Psi
= 
\label{eq:ErwinEQinCMandRELcoords}
-\frac{\hbar^2}{2M}\Delta_{\QV}^{}\Psi
-\frac{\hbar^2}{2\mu}\Delta_{\qV}^{}\Psi
 - e^2 \frac{1-e^{-\varkappa |\qV|}}{|\qV|}\Psi,
\end{eqnarray}
where now $\Psi=\Psi(t,\QV,\qV)$ (in the usual mild abuse of notation).
The separation-of-variables Ansatz $\Psi(t,\QV,\qV) = \Phi(t,\QV)\psi(t,\qV)$ splits 
this equation in two, viz.
\begin{eqnarray}
i\hbar\pddt\Phi
= \label{eq:ErwinEQinCM}
-\frac{\hbar^2}{2M}\Delta_{\QV}^{}\Phi
\end{eqnarray}
for the center-of-mass degrees of freedom, and
\begin{eqnarray}
i\hbar\pddt\psi
= \label{eq:ErwinEQinREL}
-\frac{\hbar^2}{2\mu}\Delta_{\qV}^{}\psi
 - e^2 \frac{1-e^{-\varkappa |\qV|}}{|\qV|}\psi
\end{eqnarray}
for the relative, or intrinsic, degrees of freedom.

\subsubsection{Separating off time in the intrinsic Schr\"odinger problem} \label{sec:REL}

The Ansatz $\psi(t,\qV) = e^{-iEt/\hbar}u(\qV)$ separates the time variable off from 
the position variable $\qV$ in (\ref{eq:ErwinEQinREL}), yielding the intrinsic 
eigenvalue problem 
\begin{eqnarray}
-\frac{\hbar^2}{2\mu}\Delta_{\qV}^{}u
 - e^2 \frac{1-e^{-\varkappa |\qV|}}{|\qV|} u
= \label{eq:ErwinEQstationaryREL}
E u.
\end{eqnarray}
By Theorem X.15 of Ref. \citen{RSii}, the Hamilton operator $H_\varkappa$ defined 
by the left-hand side of \refeq{eq:ErwinEQstationaryREL} is self-adjoint on $\cH^2(\Rset^3)$, 
the operator domain of $-\Delta$. By Weyl's theorem, the essential spectrum of 
$H_\varkappa$ is the positive real half-line, and by Theorem XIII.6 of Ref. \citen{RSiv}, 
$H_\varkappa$ has infinitely many eigenvalues $E<0$, each with pertinent eigenfunction 
$u\in \cH^2(\Rset^3)$.
 
We are interested in estimating these negative eigenvalues $E$.
 
\subsubsection{Separating off the angular variables in the eigenvalue problem} \label{sec:ANG}

The manifest $O(3)$ symmetry of the problem (\ref{eq:ErwinEQstationaryREL}) allows one 
to separate it completely in spherical coordinates $\qV = (r,\vartheta,\varphi)$. If 
$Y_\ell^m(\vartheta,\varphi)$ with $\ell\in\{0,1,2,3,...\}$ and $m\in\{-\ell,...,0,...,\ell\}$
denotes the standard three-dimensional spherical harmonics, satisfying
\begin{eqnarray}
-\Delta_{\Sset^2}^{} Y_\ell^m 
= \label{eq:SPHEREharm}
\ell(\ell+1)Y_\ell^m ,
\end{eqnarray}
the Ansatz $u(\qV) = R(r)Y_\ell^m(\vartheta,\varphi)$ yields the radial equation
\begin{eqnarray}\hspace{-10pt}
-\frac{\hbar^2}{2\mu}\Delta_{r}^{} R + \frac{\hbar^2\ell(\ell+1)}{2\mu r^2}R
 - e^2 \frac{1-e^{-\varkappa r}}{r} R
= \label{eq:ErwinRADIAL}
E R,\quad
\end{eqnarray}
with $rR\in L^2(\Rset_+)$, and with $\Delta_r = \frac{1}{r^2}\partial_r(r^2\partial_r\ )$.
It is obvious that $E$ does not depend on the magnetic quantum number $m$; hence,
for each given $\ell$ an eigenvalue $E$ is at least $2\ell+1$-fold degenerated.

\subsection{The radial Schr\"odinger eigenvalue problem} \label{sec:RADsolved}

\subsubsection{Switching to dimensionless physical quantities} \label{sec:dimLESS}

We now rewrite the radial Schr\"odinger equation \refeq{eq:ErwinRADIAL} in a 
dimensionless manner. The (reduced) Compton wave length of the electron, 
$\lambdaC = \frac{\hbar}{\mEL c} (\approx 3.86 \times 10^{-13}\mbox{m})$,
will serve as dimensional reference length, and the rest energy of the electron, 
$\mEL c^2 (\approx 511 \mbox{keV})$, multiplied by $(1+\mEL/\mPR)$, with 
$\mEL/\mPR\approx 1/1836$, will serve as dimensional reference energy.
Thus, in \refeq{eq:ErwinRADIAL} we make the following replacements, 
\begin{eqnarray}
& r\mapsto \lambdaC r,
\\
& \varkappa \mapsto \lambdaC^{-1}\varkappa,
\\
& E\mapsto (1+\mEL/\mPR)\mEL c^2 E,
 \label{eq:dimLESSquantities}
\end{eqnarray}
and obtain the dimensionless radial Schr\"odinger equation
\begin{eqnarray}
-\frac{1}{2}\Delta_{r}^{} R + \frac{\ell(\ell+1)}{2 r^2}R
 - \alpha \frac{1-e^{-\varkappa r}}{r} R
= \label{eq:ErwinRADIALdimLESS}
E R,
\end{eqnarray}
with
\begin{eqnarray}
\alpha = \frac{\alphaS}{1+\frac{\mEL}{\mPR}},
\label{eq:ALPHA}
\end{eqnarray}
where $\alphaS = \frac{e^2}{\hbar c} \approx \frac{1}{137.036}$ is Sommerfeld's 
fine structure constant.

\subsubsection{Rigorous Results} \label{sec:RIGOR}

In the limit $\varkappa\to\infty$, we obtain the Bohr spectrum, i.e. for each 
$\ell\in\{0,1,2,...\}$ we have
\begin{eqnarray}
E_{\ell,n}(\infty) = -\frac12 \frac{\alpha^2}{n^2}, \ n > \ell.
\label{eq:BOHRspec}
\end{eqnarray}
Note that this is equivalent to the usual textbook presentation of the Bohr energies as
$E^{\mbox{\tiny{Bohr}}}_n = -\frac12 \frac{\alpha^2}{n^2}, \ n \in\Nset$, with 
$E^{\mbox{\tiny{Bohr}}}_n$ occuring $n^2$ times; namely, \emph{given} $n$, the 
eigenvalue $E^{\mbox{\tiny{Bohr}}}_n$ occurs for each angular momentum quantum number 
$\ell$ satisfying $0\leq \ell < n$, and \emph{given} also such $\ell$, it occurs for 
each magnetic quantum number $m\in\{-\ell,...,0,...,\ell\}$. In the same limit, we 
have $\lim_{\varkappa\to\infty}R^{(\varkappa)}_{\ell,n}(r) =: R^{(\infty)}_{\ell,n}(r)$, 
where $R^{(\infty)}_{\ell,n}(r) = R^{}_{n,\ell}(r)$ is the conventional normalized radial 
Schr\"odinger eigenfunction of hydrogen with Coulomb interaction between electron and 
proton, i.e. 
\begin{eqnarray}
 R_{\ell,n}^{(\infty)}(r)  = \label{eq:RellnCOULOMB}
 \sqrt{\frac{(n-1-\ell)!}{2n(n+\ell)!}}\;
 e^{-\alpha r/n}\left(\frac{2\alpha}{n}r\right)^\ell 
 L_{n-1-\ell}^{(2\ell+1)}\left(\frac{2\alpha}{n}r\right)
\left(\frac{2\alpha}{n}\right)^{\frac32} ,
\end{eqnarray}
where $L_\nu^{(\kappa)}(\xi)$ with $\nu\in\{0,1,2,...\}$ and $\kappa\in\Rset_+$ is the 
associated Laguerre polynomial as defined in Ref. \citen{AS}, with generating function
\begin{eqnarray}
\sum_{\nu=0}^\infty t^\nu L_\nu^{(\kappa)}(\xi) = 
\frac{e^{-\frac{t\xi}{1-t}}}{(1-t)^{\kappa +1}}.
\end{eqnarray}

For every $\ell\in\{0,1,2,...\}$ there is a countable set of eigenvalues 
$E_{\ell,n}(\varkappa),\ n\in\Nset\ (n>\ell)$, with $E_{\ell,n}(\varkappa)<0$, and 
with radial eigenfunctions $R_{\ell,n}^{(\varkappa)}(r)$ satisfying $rR\in L^2(\Rset_+)$.
Those eigenvalues can be estimated as follows.

Since the BLTP pair interaction energy differs by a positive term 
$\alpha e^{-\varkappa r}/r$ from the usual Coulomb pair interaction energy, it 
follows that the Bohr energies are lower bounds to the BLTP-hydrogen energies, i.e.
\begin{eqnarray}
\forall\; \ell, n > \ell:\ E_{\ell,n}(\varkappa) > -\frac12 \frac{\alpha^2}{n^2}.
\label{eq:BOHRspecISlowerBOUND}
\end{eqnarray}

By the Rayleigh--Ritz variational principle, we also have a rigorous upper bound on the 
BLTP-hydrogen energies, obtained by adding $\alpha \langle e^{-\varkappa r}/
r\rangle_{\!\ell,n}^{}$ to the Bohr energies, where the expected values are 
computed with the $R^{(\infty)}_{\ell,n}(r)$. For general $\ell\in\{0,1,2,...\}$ and 
$n\in\Nset$ with $n> \ell$, we have 
\begin{eqnarray}
 \left\langle \dfrac{e^{-\varkappa r}}{r}\right\rangle_{\!\!\ell,n}^{} 
:=& \label{eq:BLTPpotAVEdef}
{\displaystyle \int_0^\infty} \dfrac{e^{-\varkappa r}}{r} R_{\ell,n}^{(\infty)}(r)^2 r^2 \drm{r} 
\\
 =&
{\displaystyle \int_0^\infty} {e^{-\varkappa r}} R_{\ell,n}^{(\infty)}(r)^2 r \drm{r} .
\nonumber
\end{eqnarray}
Recalling \refeq{eq:RellnCOULOMB}, setting $\xi = \frac{2\alpha}{n} r$, and invoking the 
generating function of the associated Laguerre polynomials, these expected values can be 
computed explicitly, viz.
\begin{eqnarray}
\left\langle \dfrac{e^{-\varkappa r}}{r}\right\rangle_{\!\!\ell,n}^{} \!\!\!
= \label{eq:BLTPpotAVEevalA}
 \alpha C_{\ell,n} \left( \! \! \dfrac{\partial^2}{\partial t\partial s} \! \! \right)^{\!\!n-1-\ell}
F(t,s)\Big|_{t=0\, \&\, s=0} 
\end{eqnarray}
with
\begin{eqnarray}
C_{\ell,n}
= \label{eq:Cnell}
\frac{1}{n^2(n+\ell)!(n-1-\ell)!} 
\end{eqnarray}
and
\begin{eqnarray}
F(t,s) & = 
{\displaystyle \int_0^\infty}  
\!\!\! e^{- \frac{\varkappa n}{2\alpha} \xi } \xi^{2\ell}  e^{-\xi}
\frac{e^{-\frac{t\xi}{1-t}}}{(1-t)^{2\ell+2}}
\frac{e^{-\frac{s\xi}{1-s}}}{(1-s)^{2\ell+2}}
\xi \drm{\xi}  \\
&= \frac{(2\ell+1)!}{ \Big[ \! (1 \! - \! t ) ( 1 \! - \! s) \Big( \! \big( 1+\frac{\varkappa n}{2\alpha} \big) 
\! + \! \tfrac{t}{1 \! - \! t} \! + \! \tfrac{s}{1 \! - \! s} \Big) \! \Big] ^{2\ell + 2}}. \qquad
\end{eqnarray}
 For small $n-1 - \ell$, \refeq{eq:BLTPpotAVEevalA} is readily evaluated. 
In particular, when $\ell=n-1$ we have 
\begin{eqnarray}
\alpha \left\langle \frac{e^{-\varkappa r}}{r}\right\rangle_{\!\!n-1,n}^{} 
=
\frac{\alpha^2}{n^2}\frac{1}{\left(\frac{\varkappa n}{2\alpha}+1 \right)^{2n}}, 
\qquad n\in\Nset.
\label{eq:firstORDERcorrectionELLnMINone}
\end{eqnarray}
Thus, for the energy eigenvalues $E_{n-1,n}(\varkappa)$ we have the rigorous upper 
bound 
\begin{eqnarray}
\ E_{n-1,n}(\varkappa)
 < - \frac12\frac{\alpha^2}{n^2}\left( 
1 -  2
\frac{
\left(\frac{2\alpha}{\varkappa n}\right)^{2n}
}{
\left(1+ \frac{2\alpha}{\varkappa n}\right)^{2n}
}
\right)
\label{eq:EnMinONEnUPPERbound}
\end{eqnarray}
for all $n\in\Nset$; setting $n=1$ yields an upper bound on the ground state energy. 

\begin{rem}
Using the Rayleigh-Ritz method, Cuzinatto et al. \cite{Cuzinattoetal2011} found the slightly weaker 
upper bound for the ground state energy
\begin{eqnarray}
E_{0,1}(\varkappa) < 
- \frac{\alpha  ^2}{2} \left( 1 -  2 \Big(  \frac{2 \alpha}{\varkappa} \Big)^2 \right).
\end{eqnarray}
\end{rem}

For general $0\leq \ell<n$, we can pull out $(1-t)$ $(1-s)(1+\frac{\varkappa n}{2\alpha})$
from under the fraction at the right-hand side of \refeq{eq:BLTPpotAVEevalA} and Maclaurin-expand 
${1}/{[1+x]^{2\ell+2}}$ with $x = \left(\frac{t}{1-t} + 
\frac{s}{1-s}\right)\frac{1}{1+\frac{n\varkappa}{2\alpha}}$
to find an asymptotic expansion of $\alpha 
\left\langle \frac{e^{-\varkappa r}}{r}\right\rangle_{\!\!\ell,n}^{}$
in powers of ${1}/{[1+\frac{\varkappa n}{2\alpha}}]$ greater or equal 
than ${}^{2\ell+2}$. In particular, with 
\begin{eqnarray}
 \left( \dfrac{\partial^2}{\partial t\partial s} \right)^{\!\!n-1-\ell}
\dfrac{1}{[(1-t)(1-s)]^{2\ell+2}}\biggl|_{t=0\, \&\, s=0}\biggr.   =& 
\\
\nonumber
 \left[ \! \left(\dfrac{\partial}{\partial t}\right)^{\!\!n-1-\ell}_{\! t=0} \!
\dfrac{1}{(1-t)^{2\ell+2}} \right]^2 \biggl|_{t=0}\biggr.  = & \dfrac{(n+\ell)!^2}{(2\ell+1)!^2}
\end{eqnarray}
we find for the energy eigenvalues $E_{\ell,n}(\varkappa)$ the rigorous upper bound 
\begin{eqnarray}
 E_{\ell,n}(\varkappa)
 < 
\label{eq:EellnUPPERbound}
- \frac12\frac{\alpha^2}{n^2}\left[ 1 - 
2
\left( \begin{array}{c}
n + \ell
\\
2 \ell + 1 
\end{array} \right)
\frac{\left(\frac{2\alpha}{\varkappa n}\right)^{2\ell+2}}{\left(1+ 
\frac{2\alpha}{\varkappa n}\right)^{2\ell+2}}
+\mathcal{O}\left(\left(\frac{2\alpha}{\varkappa n}\right)^{2\ell+3}\right)
\right];
\end{eqnarray}
for all $\ell \in \{0,1,2,...\}$ and $n> \ell$.

\begin{rem}
We could improve these upper bounds by replacing $\alpha$ in the hydrogen eigenfunctions 
for Couloumb interactions, see (\ref{eq:RellnCOULOMB}), with a parameter $\tilde\alpha$ 
(say), then compute the optimal $\tilde\alpha$ which minimizes the upper bound 
$ E_{\ell,n}(\varkappa)\leq \langle H \rangle(\tilde\alpha)$, where $H$ is given 
by the operator at l.h.s. (\ref{eq:ErwinRADIALdimLESS}). However, the improvement 
shows only at higher order than the order $(\alpha/\varkappa n)^{2\ell+2}$ term in 
the square bracketed expression of our upper bound (\ref{eq:EellnUPPERbound}).
\end{rem}

\begin{rem}\label{rem:alpha}
Setting $\varkappa = 2/\alphaS\approx 2/\alpha$ (this value results when the 
electrostatic field energy of a point charge equals the empirical rest mass of the 
electron), our upper bound (\ref{eq:EellnUPPERbound}) on $E_{\ell,n}(\varkappa)$ 
implies that the BLTP correction to the Bohr spectrum is of higher order in $\alphaS$ 
than the relativistic correction coming from Dirac's equation (with Coulomb interaction). 
\end{rem}

\subsubsection{Numerical  Results} \label{sec:NUM}
 
In this subsection we complement the estimates given in the preceding section with a 
few numerical results. To that end we have solved the dimensionless Schr\"odinger 
equation (\ref{eq:ErwinRADIALdimLESS}) numerically with the standard ODE solver of 
MATHEMATICA, using the shooting method. We have done this for a selection of values of 
the Bopp length $\varkappa ^{-1}$ of the BLTP theory, and quantum numbers $\ell$ and $n$. 

\begin{figure}[H]
\includegraphics[width=11.25cm]{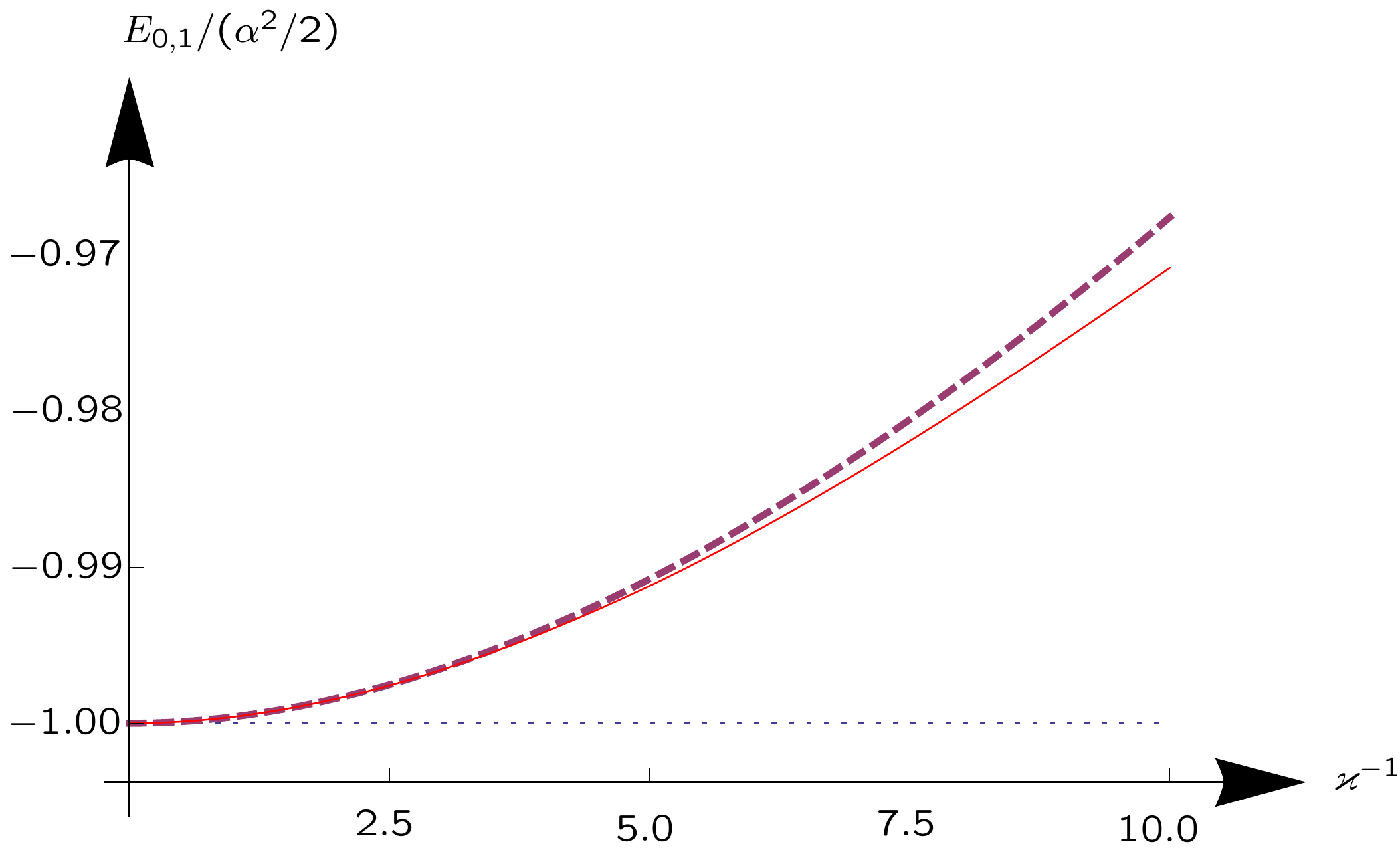}\vspace{-.5truecm}   
\caption{
The energy level $E_{\ell,n}(\varkappa)$ of the ground state ($\ell =0$, $n=1$) plotted 
against the Bopp length $\varkappa ^{-1} \in (0,10)$. The upper bound 
(\protect\ref{eq:EellnUPPERbound}) and the lower bound (\protect\ref{eq:BOHRspecISlowerBOUND}) 
are shown also (as a dashed line and as a dotted line, respectively). \label{fig:E10}
}
\includegraphics[width =11.25cm]{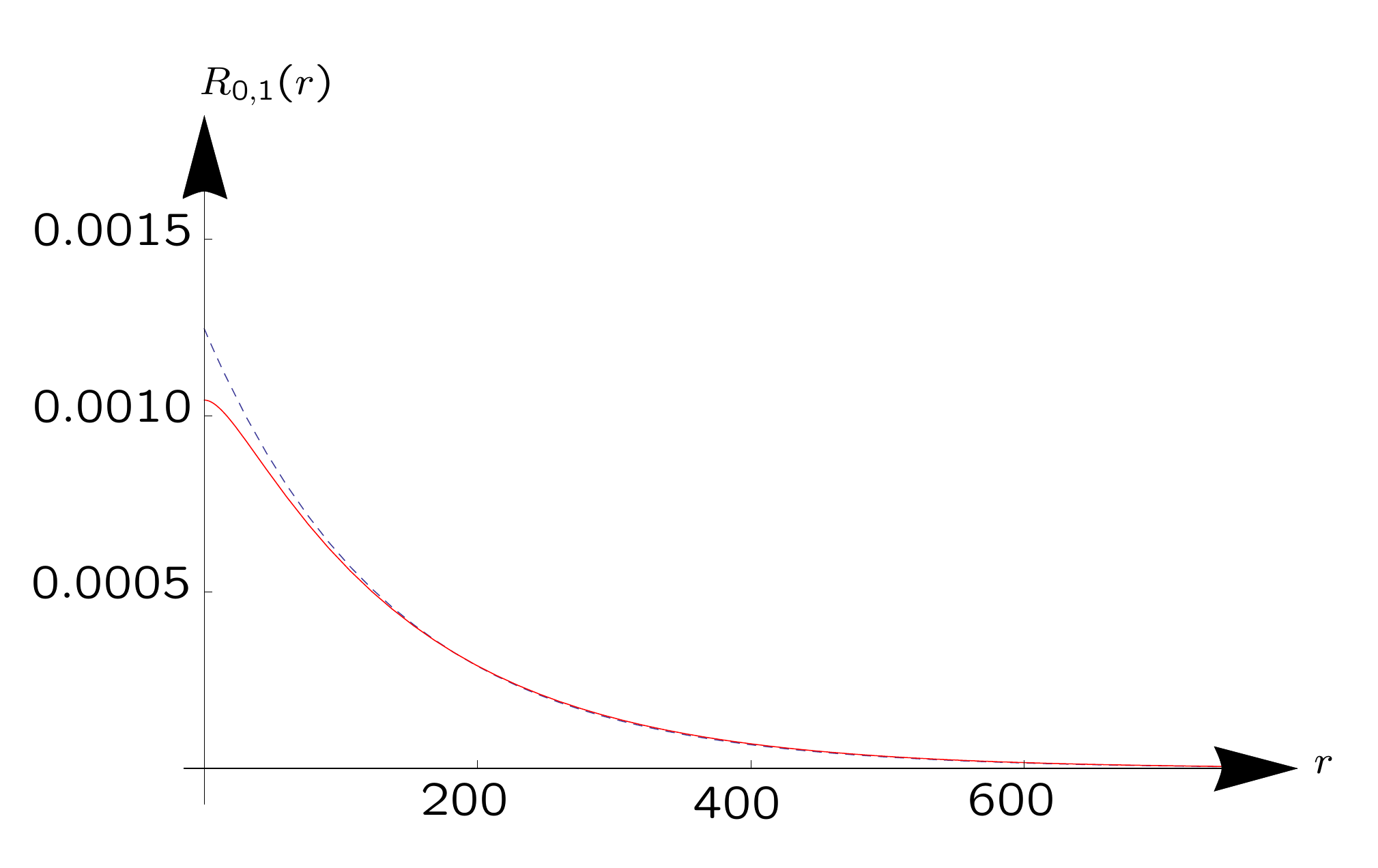}   \vspace{-.5truecm}
\caption{
Radial wave function $R^{(\varkappa)}_{\ell,n}(r)$ for the ground state 
($\ell =0$, $n=1$) with $\varkappa ^{-1} =10$ (solid) and, for the sake of 
comparison, with $\varkappa ^{-1} =0$ (dashed).   
\label{fig:R10}}
\end{figure}

Here first are the results for the ground state ($\ell =0$, $n=1$). Fixing the initial 
conditions $u(0)=0$ and $u'(0)=1$ for the function $u(r)=r R(r)$ and varying $E$, 
we have shot for the solution with $u(r) \to 0$ for $r \to \infty$ that has no zeros 
in the interval $0 < r < \infty$, then normalized $u$ afterwards. This we did for 
several values of $\varkappa ^{-1}$ in the interval $0 < \varkappa ^{-1} < 10$, then 
interpolated the results over this interval, see Fig.~\ref{fig:E10}. 

The numerically determined radial ground state wave function is shown in 
Fig.~\ref{fig:R10}. For this plot we have chosen the unrealistically high value 
of $\varkappa^{-1} = 10$, because the resulting graphs of $R_{0,1}^{(\varkappa)}(r)$ 
for $\varkappa^{-1}\ll 10$ become indistinguishable by the naked eye from the Coulomb 
case $\varkappa^{-1} =0$. We see that a non-zero $\varkappa^{-1}$ results in a diminished 
probability density near $r=0$, compared to the case $\varkappa^{-1}=0$, in agreement 
with the fact that the BLTP law of the vacuum weakens the electrical attraction at 
short distances compared to the Coulomb law. With increasing $r$ the radial wave 
function for the ground state falls monotonically to zero. Barely indicated in 
Fig.~\ref{fig:R10}: $R_{0,1}^{(\varkappa)}(r)$ approaches the Coulombic radial wave 
function, $R_{0,1}^{(\infty)}(r)$, from above as $r$ grows large. 

\begin{figure}[H]
\includegraphics[width  =11cm]{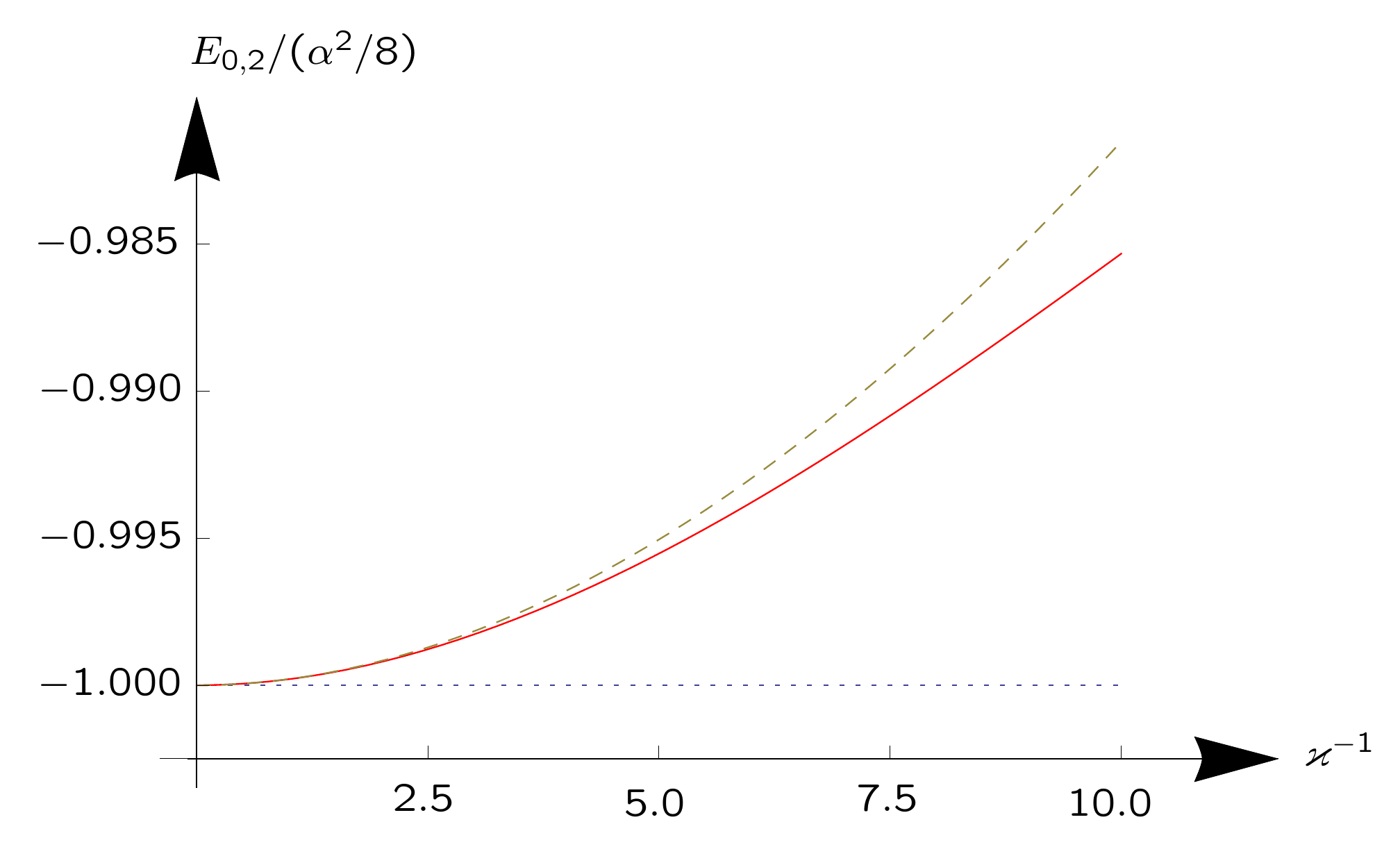}   
\caption{
 The energy level $E_{\ell,n}(\varkappa)$ for $\ell =0$ \&\ $n=2$ versus $\varkappa ^{-1}$. 
 Upper bound (\protect\ref{eq:EellnUPPERbound}) and lower bound (\protect\ref{eq:BOHRspecISlowerBOUND}) 
are shown as a dashed line and as a dotted line, respectively. \label{fig:E20}}
\includegraphics[width  =11cm]{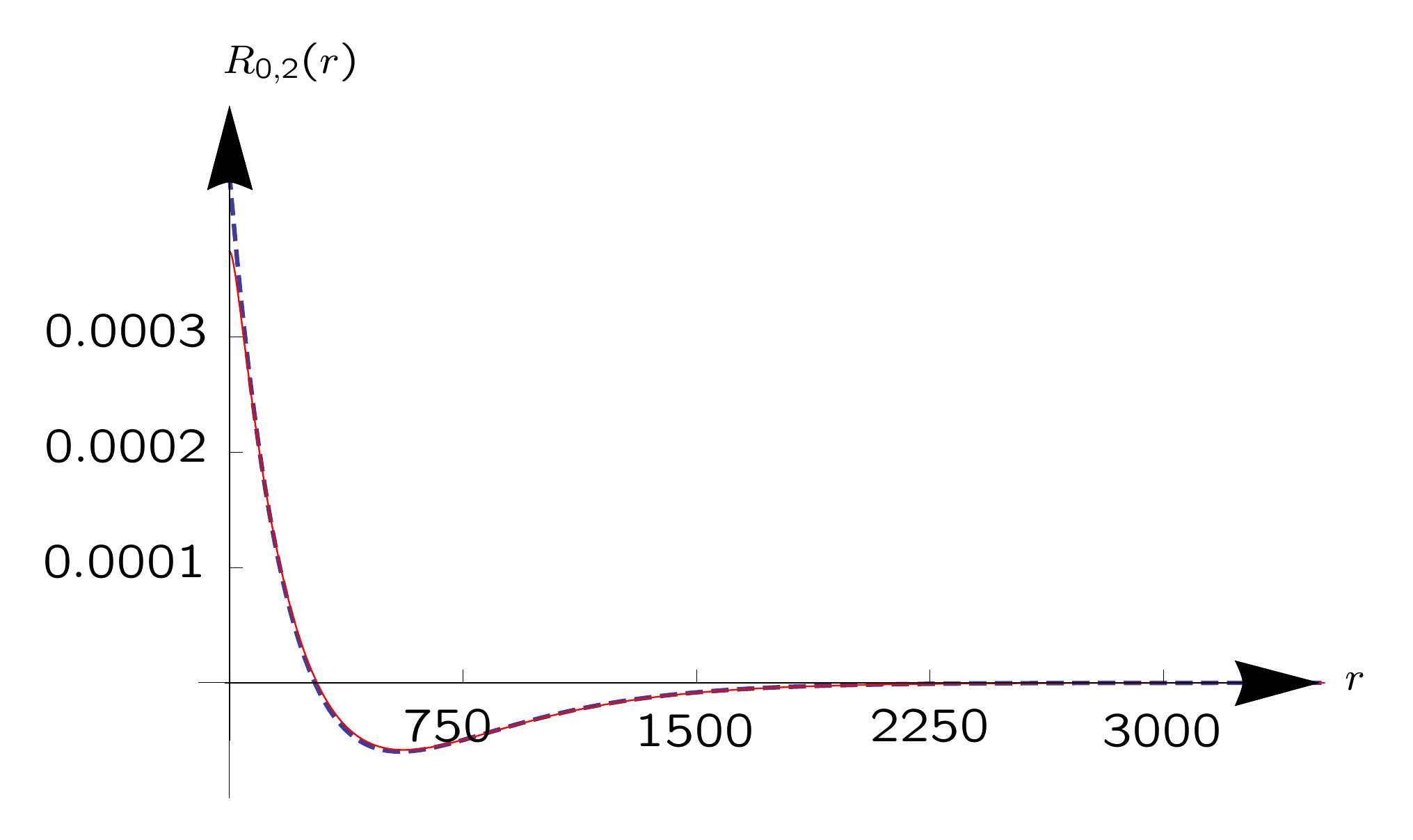}   
\caption{
Radial wave function $R^{(\varkappa)}_{\ell,n}(r)$ for $\ell =0$ \&\ $n=2$ with 
$\varkappa ^{-1} =10$ (solid) and $\varkappa ^{-1} =0$ (dashed).
 The two curves are optically barely distinguishable (near the origin). \label{fig:R20}}
\end{figure}

We now turn to the case $n=2$. In the case $\ell =0$ we use the same numerical method 
as for the ground state. The results are shown in Figs. \ref{fig:E20} and \ref{fig:R20}.

\begin{figure}[H]
\includegraphics[width =11cm]{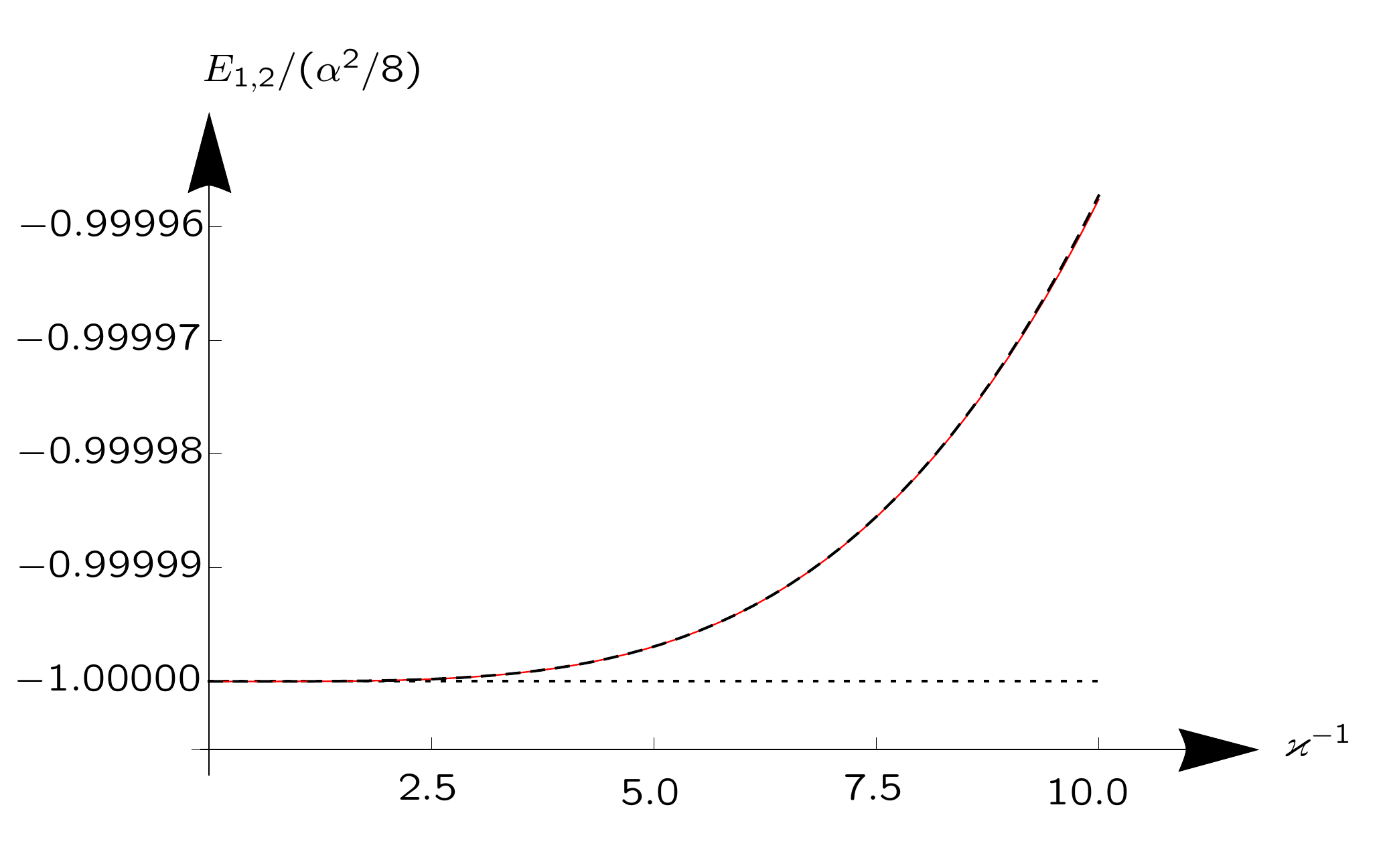}   
\caption{
The energy level $E_{\ell,n}(\varkappa)$ for $\ell =1$ \&\ $n=2$ versus $\varkappa ^{-1}$. 
The upper bound (\protect\ref{eq:EellnUPPERbound}) (dashed) and the lower 
bound (\protect\ref{eq:BOHRspecISlowerBOUND}) 
(dotted) are shown as well. \label{fig:E21}}
\includegraphics[width  =11cm]{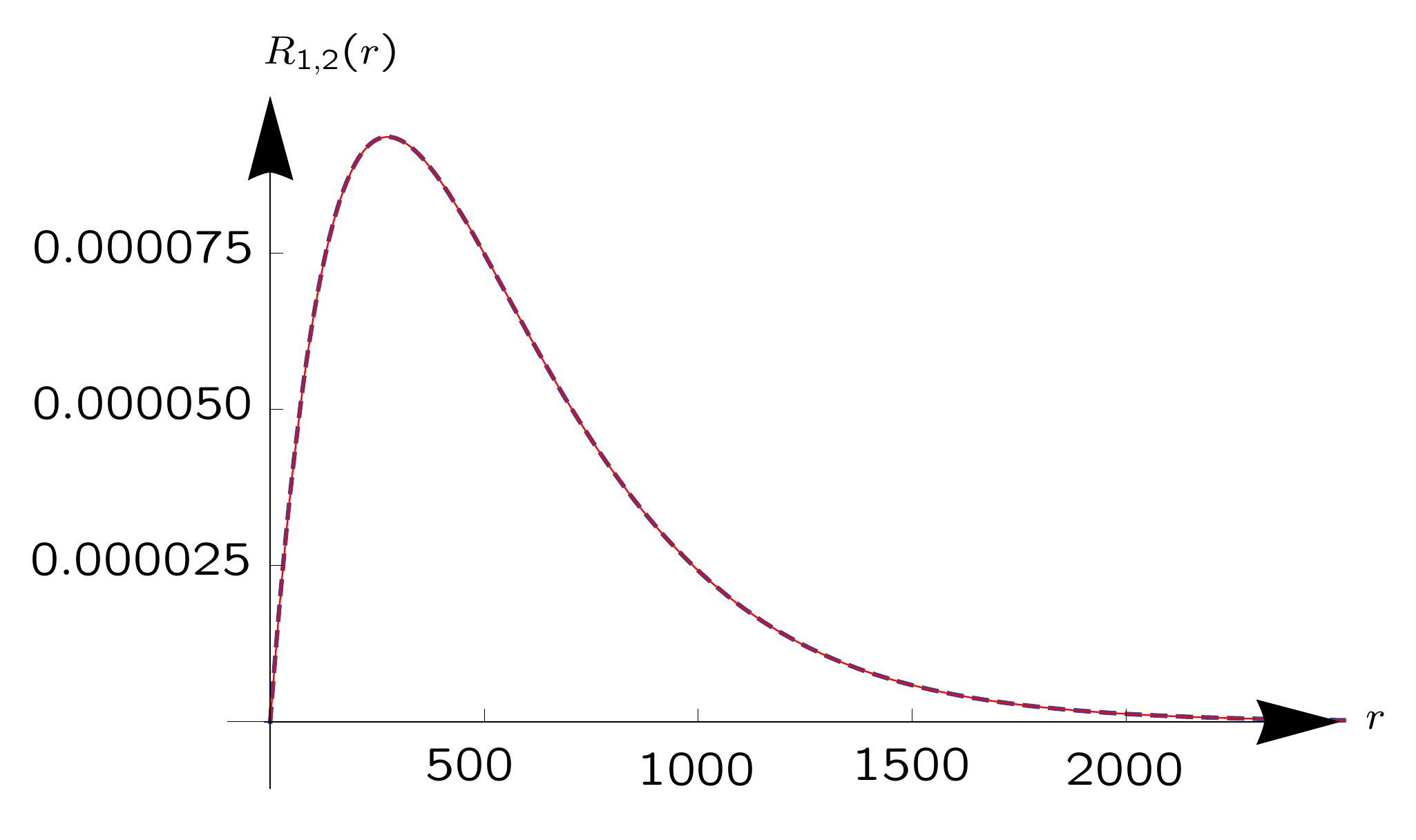}   
\caption{
Radial wave function $R^{(\varkappa)}_{\ell,n}(r)$ for $\ell =1$ \&\ $n=2$ with 
$\varkappa ^{-1} =10$ (solid) and $\varkappa ^{-1} =0$ (dashed).
 The two curves are virtually indistinguishable by the naked eye. \label{fig:R21}}
\end{figure}

To deal with the case $\ell =1$ is more cumbersome. As the centrifugal potential in 
the Schr\"odinger equation (\ref{eq:ErwinRADIALdimLESS}) is singular at $r=0$, the 
equation cannot be numerically solved by giving initial conditions at $r=0$. 
Therefore we give initial conditions at $r=10^{-7}$ and shoot for the solution 
with $u(r) =0$ and $u(r) \to 0$ for $r \to \infty$  by varying the energy and the 
initial conditions. The results are shown in Figs. \ref{fig:E21} and \ref{fig:R21}. 

Figures \ref{fig:E10}, \ref{fig:E20}, and \ref{fig:E21} reveal that the upper 
bound (\protect\ref{eq:EellnUPPERbound}) is a very good approximation for the numerically 
computed energy for all values of $\varkappa ^{-1}\leq 5$; for $E_{1,2}(\varkappa)$ this 
is even true over the entire interval $0 < \varkappa ^{-1} < 10$. This tendency of 
increased accuracy with increasing $\ell$ is directly reflected in our upper bound 
(\protect\ref{eq:EellnUPPERbound}), which becomes closer to the Coulomb eigenvalues 
with increasing $\ell$.

Figures \ref{fig:E20} and \ref{fig:E21} also reveal that $E_{1,2}(\varkappa)$ changes 
much less with $\varkappa ^{-1}$ than $E_{0,2}(\varkappa)$. As a consequence, the 
well-known fact that the singlet state ($\ell =0$) and the triplet state ($\ell =1$)
have the same Schr\"odinger energy when $\varkappa^{-1}=0$ no longer holds 
when $\varkappa^{-1}>0$; more precisely, $E_{0,2} ( \varkappa ) > E_{1,2} 
(\varkappa )$ for $\varkappa ^{-1} \neq 0$.

\begin{figure}
\includegraphics[width  =11cm]{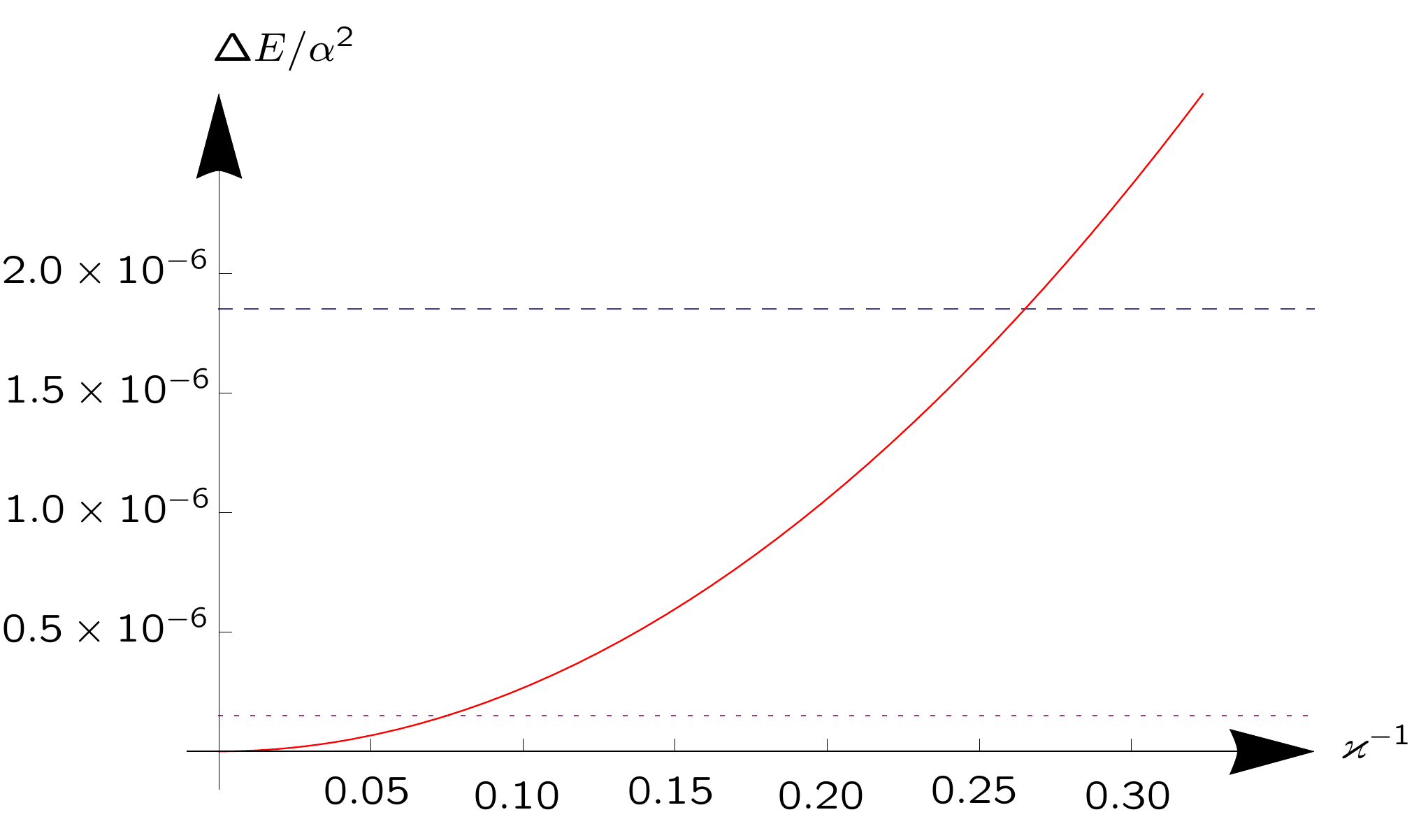}   
\caption{
BLTP splitting (\protect\ref{eq:DeltaE}) of the Lyman-$\alpha$ 
transition in comparison to the fine-structure splitting 
$\Delta E _{\mathrm{fine}} \approx 1.85 \times 10^{-6} \alpha ^2$ (dashed line)
and the Lamb shift $\Delta E _{\mathrm{Lamb}} \approx 1.50 \times 10^{-7} \alpha ^2$
(dotted line). \label{fig:Lya1}}
\end{figure}

The fact that $E_{0,2} ( \varkappa ) > E_{1,2} (\varkappa )$ for $\varkappa^{-1} \neq 0$ 
means that the BLTP interaction gives rise to
a $\varkappa$-dependent splitting of the Lyman-$\alpha$ line (i.e., of the spectral line 
that corresponds to the transition from the $(n=2)$ level to the $(n=1)$ level) 
already with the non-relativistic Schr\"odinger Hamiltonian.
 From our numerically computed energy values we can calculate 
the separation 
\begin{eqnarray}\label{eq:DeltaE}
 \Delta E ( \varkappa )
:= 
 \big(E_{0,2}(\varkappa)-E_{0,1}(\varkappa)\big) - 
\big(E_{1,2}(\varkappa)-E_{0,1}(\varkappa)\big)
= 
 E_{0,2}(\varkappa)-E_{1,2}(\varkappa) 
\end{eqnarray}
of these two Lyman-$\alpha$ components.  
In Fig.~\ref{fig:Lya1} the result is plotted versus $\varkappa ^{-1}$.
Both energy differences $E_{1,2}(\varkappa)-E_{0,1}(\varkappa)$ 
and  $E_{0,2}(\varkappa)-E_{0,1}(\varkappa)$ are measurable: The first one corresponds 
to the two dominant components of the Lyman-$\alpha$ line coming from the 
by far most likely spontaneous transitions from the $n=2$ to the $n=1$ level, namely 
from the $2{}^2\!P_{\frac12}$ state to the $1{}^2\!S_{\frac12}$ state and from the 
$2{}^2\!P_{\frac32}$ state to the $1{}^2\!S_{\frac12}$ state. These transitions have 
been routinely observed for more than a century. The second one is more difficult to 
observe; the single-photon transition from the $2S$ to the $1S$ state is forbidden for 
spinless particles in spherically symmetric potentials, i.e., it can be realised as a 
single-photon transition only if the electron spin is taken into account, and even then 
it has a very low transition probability. However, it has been observed and measured with
high accuracy as a two-photon transition, see Parthey et al. \cite{PartheyEtAl2011}.

Except for a subtlety which we have to discuss (see next section), the default criterion 
for acceptable $\varkappa$-modifications of the theoretical hydrogen spectrum would be 
that the differences are not larger than the measurement uncertainty of the empirical 
spectra. As a figure of merit for the latter we may use the fine structure splitting of the 
Lyman-$\alpha$ line which is usually explained as a relativistic quantum-mechanical 
effect. It has been calculated in the Born--Oppenheimer approximation\footnote{To go 
beyond the Born--Oppenheimer approximation one usually starts from the Pauli spectrum 
of hydrogen in the center-of-mass system (as we have done for the Schr\"odinger spectrum)
and computes relativistic corrections to it in powers of $\alphaS$ and $\ln\alphaS$.} by replacing the 
Schr{\"o}dinger equation with the Dirac equation for an electron in the Coulomb field 
of an infinitely massive proton. This Dirac energy eigenvalue spectrum is identical 
with the Sommerfeld fine-structure formula except for its angular momentum labelling; 
see Ref. \citen{Keppeler} for an illuminating analysis. The fine-structure splitting of the 
Lyman-$\alpha$ line is essentially due to the different spin-orbit coupling strength
of the $2\,{}^2\!S_{\frac12}$/$2\,{}^2\!P_{\frac12}$ and $2\,{}^2\!P_{\frac32}$ states; recall 
that the energies of the $2\,{}^2\!S_{\frac12}$ and $2\,{}^2\!P_{\frac12}$ states coincide 
in this Dirac spectrum. The numerical result, which is well confirmed by observation,
for the two transition frequencies of the Lyman-$\alpha$ line is $\nu _1 \approx 2.466060 
\times 10^{15} \mathrm{Hz}$ and $\nu _2 \approx 2.466072 \times 10^{15} \mathrm{Hz}$ 
which, in our units, corresponds to an energy difference of 
$\Delta E _{\mathrm{fine}} \approx 1.85 \times 10^{-6} \alpha ^2$. This value
is marked in Figure \ref{fig:Lya1} as a dashed line. The fact that a BLTP
modification of the hydrogen spectrum has not been observed so far requires
that the $\Delta E ( \varkappa )$ from (\ref{eq:DeltaE}) must be small in comparison
to $\Delta E _{\mathrm{fine}}$. From Figure \ref{fig:Lya1} we read that this 
criterion requires $\varkappa ^{-1}$ to be significantly smaller than $0.25$.

This is the upper bound we get from evaluating the difference of the two
Lyman-$\alpha$ transition energies as given in the middle line of (\ref{eq:DeltaE}).
As an alternative, we may view $\Delta E (\varkappa )$ as the direct transition
energy from the ($n=2, \ell =0$) level to the ($n=2, \ell=1$) level. In the
standard theory with $\varkappa ^{-1} =0$ this transition is possible because
the energy of the $2^2S_{1/2}$ state differs from the $2^2P_{1/2}$ state by the Lamb
shift. The latter has been measured by Lamb and Retherford \cite{LambRetherford1947}
in 1947 and theoretically calculated, on the basis of quantum electrodynamics, 
by Bethe \cite{Bethe1947} in the same year, with the result that the transition 
frequency is about 1 GHz. In our dimensionless units this corresponds to a transition 
energy of $\Delta E_{\mathrm{Lamb}} \approx 1.50 \times 10^{-7} \alpha ^2$, see 
the dotted line in Figure \ref{fig:Lya1}. As the observed value of the Lamb
shift is in agreement with the theoretically predicted one, to within the
present measuring accuracy, we may conclude that $\Delta E (\varkappa )$ must
be significantly smaller than $\Delta E_{\mathrm{Lamb}}$. From Figure \ref{fig:Lya1}
we read that this requires $\varkappa ^{-1}$ to be significantly smaller than 
$0.075$ which is a slightly more restrictive bound than the one from comparison
with the fine structure splitting of the Lyman-$\alpha$ line. 

\begin{figure}
\includegraphics[width  =11cm]{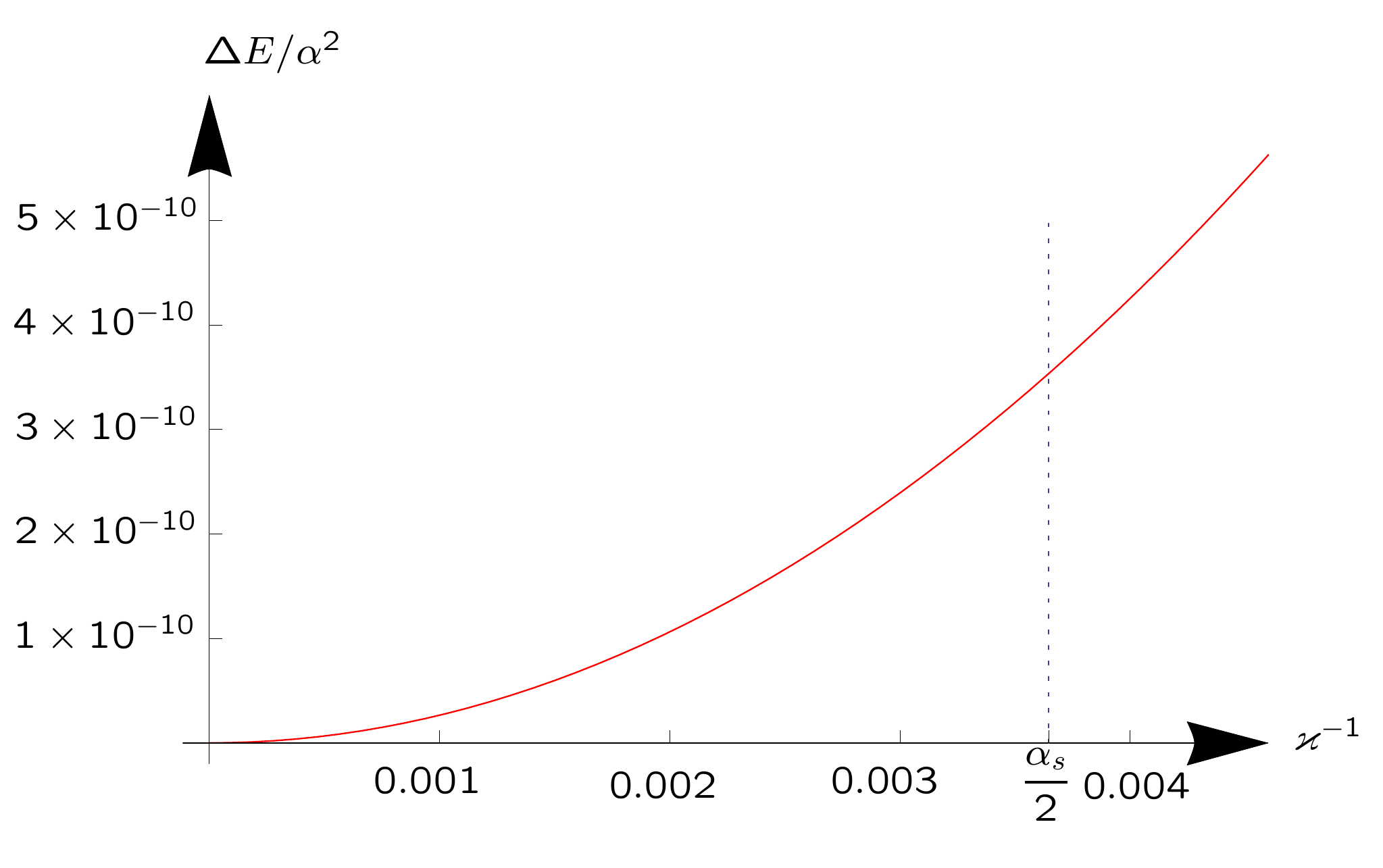}   
\caption{\small
BLTP splitting (\protect\ref{eq:DeltaE}) of the Lyman-$\alpha$ 
transition for $\varkappa ^{-1} = \alphaS /2 \approx 1/274
\approx 0.0036$. \label{fig:Lya2}}
\end{figure}

We also numerically studied the spectrum for the much smaller and theoretically 
interesting value of $\varkappa ^{-1} = \alpha _S /2$ mentioned in Remark \ref{rem:alpha}.
For this value of $\varkappa ^{-1}$ we find a splitting of the 
Lyman-$\alpha$ line of $\Delta E \approx 3.5 \times 10 ^{-10} \alpha ^2$, see 
Fig.~\ref{fig:Lya2}. This splitting is almost four orders of magnitude smaller than 
the empirical fine-structure splitting and about three orders of magnitude
smaller than the Lamb shift, which would seem to be acceptable. 
Whether this is so will be analyzed in the next section.

\section{Spectroscopic constraints on $\varkappa$} \label{sec:Viability} 
 
In the preceding section we have calculated the BLTP splitting of the 
Lyman-$\alpha$ transition (\ref{eq:DeltaE}) in the (spinless) Schr\"odinger spectrum.
This splitting is caused entirely by the fact that an electron in the 
$\ell=0, n=2$ state is more likely to be close to the nucleus than an electron 
in the $\ell=1, n=2$ state, and the influence of nonzero $\varkappa^{-1}$ is 
largest at short distance. Thus, if one now proceeds along the lines of the usual 
relativistic perturbation argument, i.e. switching to the two-body Pauli equation 
to take electron and proton spin into 
account and then adding the various relativistic correction terms --- in particular 
the spin-orbit term ---, then this nonrelativistic BLTP-induced splitting would modify 
the usual Lyman-$\alpha$ fine structure. All three levels, $2\,{}^2\!S_{\frac12}$, 
$2\,{}^2\!P_{\frac12}$, and $2\,{}^2\!P_{\frac32}$ should be lifted a little bit, yet the 
$2\,{}^2\!S_{\frac12}$ level much more than $2\,{}^2\!P_{\frac12}$ and $2\,{}^2\!P_{\frac32}$ 
levels, the latter presumably still showing essentially the same fine-structure splitting 
as for the Coulomb case. By fine-tuning $\varkappa$ one may be able to achieve that the 
$2\,{}^2\!S_{\frac12}$ and $2\,{}^2\!P_{\frac32}$ levels coincide, so that energetically 
again there would be only two different $n=2$ fine-structure levels. Such a fine-tuning 
of $\varkappa$ based on just two levels would most likely be visible in a distortion of 
the $n=3$ fine structure, and thus be unacceptable. In any event, even for the fine-tuned 
$n=2$ levels their changed degeneracy would now lead to a noticeably different hyperfine 
structure. This line of argument leads to the conclusion that $\varkappa ^{-1}$ must be 
significantly smaller than $0.25$, see Fig.~\ref{fig:Lya1}.

We thus come to discuss the special value $\varkappa ^{-1} = \alphaS /2$. We already 
noted that for this $\varkappa$ value the Lyman-$\alpha$ splitting predicted by the 
Schr\"odinger eigenvalue problem with BLTP interaction, \refeq{eq:ErwinEQstationaryREL}, 
is almost four orders of magnitude smaller than the observed fine-structure splitting, 
and would thus seem acceptable. However, the precision with which the Lyman-$\alpha$ 
transition energy itself has been measured 
--- and computed perturbatively with the ``standard model of hydrogen'' --- 
is so impressive that the influence of a $\varkappa ^{-1} = \alphaS /2$ 
is still too big for being in agreement with empirical data. 

Explicitly, for comparison with experiments we concentrate on the transition from the 
$(n=2,\ell=0)$ level to the $(n=1,\ell =0)$ level of hydrogen. The transition frequency  
has been measured with the help of two-photon spectroscopy
by Parthey et al. \cite{PartheyEtAl2011} as 2466061413187035 Hz with an absolute 
uncertainty of 10 Hz. By multiplication with Planck's constant, a frequency uncertainty 
of 10 Hz corresponds to an energy uncertainty of $4.1 \times 10^{-17} \mathrm{keV}$. 
After dividing by the rest energy of the electron we find that our dimensionless 
energy $E_{0,2}-E_{0,1}$ is known with an absolute uncertainty of $\approx 8 \times 10^{-20}$. 

Even though we would need the relativistic ``BLTP hydrogen'' spectrum of a point electron 
and a finite-size proton in order to make a definitive comparison, we can still extract 
a tentative bound on $\varkappa$ by \emph{assuming} that the $\varkappa$-induced 
spectral line shifts in a relativistic model of a point electron bound to a finite-size 
proton are comparable in magnitude to those computed here with the non-relativistic 
Schr\"odinger model of a point electron bound to a point proton. Thus, we write the 
$\varkappa$-dependent theoretical Lyman-$\alpha$ transition energy as $E_{0,2} 
( \varkappa )-E_{0,1} ( \varkappa ) = E _{0,2} ( \infty ) - E_{0,1} ( \infty ) 
+ \delta E ( \varkappa )$ and demand that $\varkappa$ is large enough so that the 
$\varkappa$-induced fine structure splitting $\delta E ( \varkappa )$ can be ignored.
From our numerical calculations we find that $| \delta E ( \varkappa ) | > 
10^{-19}$ if $\varkappa ^{-1} > 3 \times 10^{-6}$. We conclude that present-day 
spectroscopic precision measurements restrict the Bopp length $\varkappa ^{-1}$ 
to values smaller than $3 \times 10^{-6}$. Recall that we give $\varkappa ^{-1}$ in 
units of the (reduced) Compton wave length $\lambdaC$ of the electron. So 
$\varkappa^{-1}$ is restricted to values at least two orders of magnitude smaller 
than the so-called classical electron radius, and thus also at least two orders of 
magnitude smaller than the empirical proton radius.

 One might suspect that finite proton size effects will invalidate our conclusion, but
this is not the case. 
 To estimate the influence of the finite proton size, consider the extreme model 
in which the proton's electric charge is uniformly distributed over a spherical shell of radius $\rPR$. 
 The BLTP interaction of such a charge distribution with a point electron is easily computed
(cf. Ref. \citen{KiePerJMP} for balls),
\begin{equation}\label{eq:VspherePTint}
V_\varkappa(r) = -\alpha\frac{\min\left(\frac{r}{\rPR},1\right)
-\frac{{e^{-\varkappa |r-\rPR|}-e^{-\varkappa (r+\rPR)}}}{2\varkappa\,\rPR}}{r};
\end{equation}
in the limit $\varkappa\to\infty$ the $\varkappa$-dependent contributions in \refeq{eq:VspherePTint} vanish and 
one obtains the familiar Coulomb interaction $ -\frac{\alpha}{r}$ for $r>\rPR$ and $-\frac{\alpha}{\rPR}$ for
$r\leq\rPR$.

 Now recall that our analytical upper bounds on the numerically computed eigenvalues turned out to be
very accurate for ``point-proton hydrogen'' when $\varkappa\, \rPR > 1$ (say). 
 The reason is easy to understand: even though the coupling constant ($\alpha$) is the same for the 
Coulomb and the Yukawa term in the BLTP interaction (cf. footnote \ref{fn:ONE}), the Yukawa term \emph{is}
a form perturbation of the Schr\"odinger operator with pure Coulomb interaction in the sense of Kato (see Ref. \citen{RSiv});
note though that this does not invoke an expansion in powers of $\varkappa^{-1}$. 
 Its contribution can be made arbitrarily small by making $\varkappa$ large enough.
 Therefore the usual first-order perturbation theory formalism (proceding as if there was a small coefficient in front
of the Yukawa term) will give the dominant correction to the Coulombic interaction spectrum. 
 For ``point proton hydrogen'' this is precisely our term \refeq{eq:BLTPpotAVEdef} which, when added to the Coulombic
eigenvalues, produces the upper bound on the BLTP-type eigenvalues. 

\begin{figure}[H]
\includegraphics[width  =10cm]{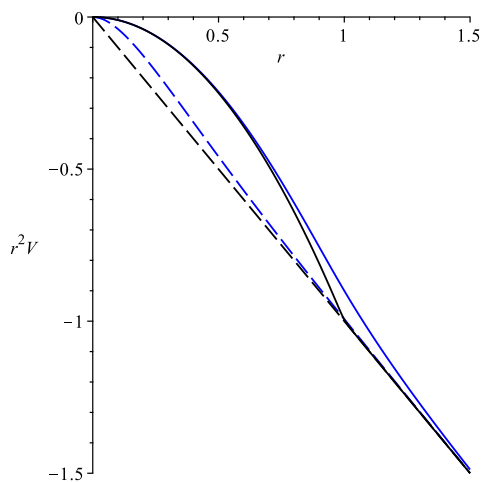} \vspace{-6truecm}
\caption{\small{\color{black}Shown is $r^2 V(r)$ (up to a constant factor) versus $r$ (in units of the proton radius $\rPR$)
 for a point electron interacting with a point proton (dashed lines), or with a spherical shell proton (continuous lines), 
once with $V$ the Coulomb (black) and once with $V$ the BLTP interaction for $\varkappa\, \rPR=5$ (blue).}
\label{fig:sphereVSpoint}}
\end{figure}

 This form perturbation argument does not rely on having a point electron 
interacting with a point proton; it applies equally well when a point electron interacts with a spherical shell proton.
 Finally, recall that the radial wave functions $R^{(\infty)}_{\ell,n}(r)$ for $n\in\{1,2\}$ vary appreciably only on 
the length scale of the Bohr radius, while on the length scale of the proton radius $\rPR$ they are essentially constant; this
is true for the ``point proton hydrogen'' as well as for ``spherical-shell proton hydrogen.''
 For $\ell =0$ and $n=1$, resp. $n=2$, these two $S$ states take non-zero values at the origin (the $n=2$ state
somewhat smaller than the $n=1$ state), whereas for $\ell = 1$ and $n=2$ this $P$ state vanishes at the origin.
 Thus to estimate the leading order effect when ``switching on $\varkappa$,'' it suffices to simply inspect the
area between the graphs of $r^2 V_\varkappa(r)$ and $r^2 V_\infty(r)$, see Fig.~\ref{fig:sphereVSpoint} 
for when $\varkappa\,\rPR = 5$.  \vspace{-1truecm}

 Fig.~\ref{fig:sphereVSpoint} suggests that replacing a point proton by a finite size proton will not 
significantly alter the effects on the spectrum caused by a change from a Coulomb to a BLTP interaction.
 We computed that the area between~the two continuous curves (obtained with the spherical shell proton model)
is almost exactly as large as the area between the two dashed curves (obtained with the point proton model), both 
when $\varkappa\,\rPR =5$ and when $\varkappa\,\rPR=100$. 
 
 We conclude that finite size effects of the proton will not likely invalidate our analysis that spectral data restrict 
$\varkappa^{-1}$ to values at least two orders of magnitude smaller than the empirical proton radius $\rPR$.

\section{Summary and Outlook} \label{sec:SummaryOutlook} 

\subsection{Summary} \label{sec:Summary} 

In the present paper we have studied the Schr\"odinger spectrum of a hydrogen atom 
when the conventional Coulomb pair energy between a point electron and a point proton 
is replaced by its Bopp--Land\'e--Thomas--Podolsky modification. We have obtained 
rigorous upper and lower bounds on the eigenvalues, and we have numerically computed a
few low lying eigenvalues as functions of the Bopp length parameter $\varkappa^{-1}$.
Based on our comparison of the theoretical curves with empirical data on the 
Lyman-$\alpha$ transition we have concluded that $\varkappa^{-1}$ has to be smaller 
than about $3 \times 10^{-6}$ (reduced) Compton wave lengths of the electron for the 
theoretical spectrum not to disagree with the experimental results. 

Converted to SI units, $\lambdaC\approx 3.86\times 10^{-13}\,\mathrm{m}$, so our 
limit $\varkappa^{-1}\lessapprox 3\times 10^{-6}\lambdaC$ means that $\varkappa^{-1}$ 
cannot be bigger than $\approx 10^{-18} \, \mathrm{m}$. Curiously, this is comparable 
to the currently available bound $R\lessapprox 10^{-18} \, \mathrm{m}$ on the size of 
the electron deduced in Ref. \citen{BrDrPRD} from experimental data and quantum 
field-theoretical considerations.

Our ``non-relativistic bound'' on $\varkappa$ also means that the old idea of 
``a purely electromagnetic origin of the electron's inertial mass,'' 
see Refs. \citen{Bopp}, \citen{landethomas}, \citen{lorentzENCYCLOP}, \citen{Abraham}, and \citen{BornA}, 
is untenable in BLTP electrodynamics, for it would require $\varkappa^{-1} =e^2/2\mEL c^2$ 
(or $\alphaS/2$ in our dimensionless units), i.e. half the so-called ``classical electron 
radius,'' in flagrant vio\-la\-tion of our non-relativistic bound on $\varkappa$. Indeed, a 
Bopp length $\varkappa^{-1} \approx 10^{-18}$m or less implies that the electrostatic 
Maxwell-BLTP field energy of a point electron is much larger than its empirical rest energy 
$\mEL c^2$. If, as assumed in all ``renormalized theories'' of the electron, the empirical 
rest mass of a physical electron is the sum of its bare rest mass and its electrostatic 
field energy, then in BLTP electrodynamics the electron has to be assigned a negative 
bare rest mass.

\newpage
\subsection{Outlook} \label{sec:Outlook} 

It should be possible to refine our study of the influence of the 
Bopp--Land\'e--Thomas--Podolsky vacuum law of electromagnetism on the 
hydrogen spectrum by replacing the Schr{\"o}dinger with the Pauli equation 
for hydrogen such that both electron and proton spin are incorporated. It 
is also possible to take the finite size of the proton into account 
in some more realistic model manner than the spherical shell proton model discussed
at the end of section \ref{sec:Viability}. Relativistic corrections should also be perturbatively 
computable. Unfortunately, a non-perturbative truly Lorentz-covariant study 
of the hydrogen spectrum, involving some version of a ``two-body Dirac operator,'' 
has been an elusive goal with the standard Maxwell vacuum law, and this is not 
going to improve by improving the vacuum law; for the time being a study of the 
Dirac spectrum of ``BLTP hydrogen'' in the Born--Oppenheimer approximation should 
be possible. We expect that such studies will only yield a refinement 
of our conclusions but no significant changes. In particular, our lower bound on 
$\varkappa$ obtained from discussing the Schr\"odinger spectrum of hydrogen
is so far removed from the theoretical value obtained by demanding a vanishing 
bare rest mass of the electron, that we expect that all these studies will suggest 
a negative bare rest mass of the electron in BLTP electrodynamics.

It is an interesting question whether other modifications of the electromagnetic 
vacuum law, for instance the Born--Infeld law \cite{BornInfeldBb}, also require a 
negative bare rest mass of the electron to be compatible with empirical spectroscopic 
data. The influence of the Born--Infeld nonlinearity on the theoretical hydrogen 
spectrum has already been investigated by various authors, see Refs.
\citen{HellerMotz}, \citen{GreinerETalA}, \citen{GreinerETalB},
\citen{CarKiePRL} and \citen{Franklin}, using various approximations, 
leading to different conclusions.\footnote{Incidentally, one discrepancy, different 
values for the $\ell=1$ Schr\"odinger eigenvalues computed with the same approximation 
to the Born--Infeld pair energy in Ref. \citen{CarKiePRL} and in Ref.
\citen{Franklin} are due to factor 2 error in the program used in 
Ref. \citen{CarKiePRL} which showed only if $\ell\neq 0$. After correcting 
this programming error, the $\ell=1$ eigenvalues came out the same as 
in Ref. \citen{Franklin}.} The road block is the formidable Born--Infeld nonlinearity, which 
in the electrostatic limit reduces to the Born nonlinearity\cite{BornA}. If the point 
charges are replaced by sufficiently smeared out charges a convergent explicit series 
expansion to solve the static problem has been constructed in 
Ref. \citen{CarKieMPAG}, but 
the algorithm does not apply to point charges. A unique two-point charge solution to 
the electrostatic Born--Infeld equations is known to exist\cite{KieMBIinCMP}, but 
the nonlinearity has so far stood in the way of finding a sufficiently \emph{accurate 
and efficient} computation of the electrostatic pair-energy of two point charges.
Once this technical obstacle has been overcome the road is paved for a systematic 
study of the Born--Infeld effects on the Schr\"odinger, Pauli, and Dirac spectra of 
hydrogen.\vspace{-.5truecm}

\section*{Acknowledgment}\vspace{-.3truecm}
We thank Shadi Tahvildar-Zadeh and Vu Hoang for helpful discussions.
VP gratefully acknowledges support from the DFG within the Research Training 
Group 1620 ``Models of Gravity.''

\newpage

\bibliographystyle{apsrev}

\begin{thebibliography}{10}

\bibitem{Bopp}
F. Bopp, 
\textit{Eine lineare Theorie des Elektrons},
Annalen Phys. (Leipzig) {\bf 430}, 345--384 (1940).

\bibitem{landethomas}\vspace{-.3truecm}
A. Land\'e A and L. H. Thomas,
\textit{Finite self-energies in radiation theory. Part II},
Phys. Rev. {\bf 60}, 514--523 (1941).

\bibitem{Podolsky}\vspace{-.3truecm}
B. Podolsky,
\textit{A generalized electrodynamics. Part I: Non-quantum},
Phys. Rev. {\bf 62}, 68--71 (1942).

\bibitem{PodolskySchwed}\vspace{-.3truecm}
B. Podolsky and P. Schwed,
\textit{A review of generalized electrodynamics},
Rev. Mod. Phys. {\bf 20}, 40--50 (1948).

\bibitem{GratusETal}\vspace{-.3truecm}
J. Gratus, V. Perlick and R. W. Tucker,
\textit{On the self-force in Bopp--Podolsky electrodynamics},
J. Phys. A: Math. Theor. {\bf 48}, 435401 (2015).

\bibitem{mikishadi}\vspace{-.3truecm}
M. K.-H. Kiessling and A. S. Tahvildar-Zadeh,
\textit{Bopp--Land{\'e}--Thomas--Podolsky electrodynamics as initial value problem},
Rutgers Univ. preprint, in preparation (2019).

\bibitem{Kvasnica1960}\vspace{-.3truecm}
J. Kvasnica,
\textit{A possible estimate of the elementary length in electromagnetic interactions},
Czech. J. Phys. {\bf 10}, 625--627 (1960).

\bibitem{Cuzinattoetal2011}\vspace{-.3truecm}
R. R. Cuzinatto, C. A. M.  de Melo, L. G. Medeiros and P. J. Pompeia,
\textit{How can one probe Podolsky electrodynamics?}
Int. J. Modern Phys. {\bf A 26}, 3641--3651 (2011).

\bibitem{Hellmann1935}\vspace{-.3truecm}
H. Hellmann,
\textit{A new approximation method in the problem of many electrons},
J. Chem. Phys. {\bf 3}, 61 (1935).

\bibitem{Hellmann1937}\vspace{-.3truecm}
H. Hellmann,
{\it Einf{\"u}hrung in die Quantenchemie},
(Franz Deuticke, Leipzig, 1937).

\bibitem{Adamowski1985}\vspace{-.3truecm}
J. Adamowski, 
\textit{Bound eigenstates for the superposition of the Coulomb and the Yukawa potentials},
Phys. Rev. {\bf A 31}, 43--50 (1985).

\bibitem{AmoreFernandez2014}\vspace{-.3truecm}
P. Amore and F. Fern{\'a}ndez,
\textit{On an approximation to the Schr{\"o}dinger equation with the Hellmann potential},
(arXiv:1411.4871v1, 2014).

\bibitem{VandenBergheFackMeyer1989}\vspace{-.3truecm}
G. Vanden~Berghe, V. Fack and H. E. De Meyer,
\textit{Numerical methods for solving radial Schr{\"o}dinger equations},
J. Comput. Appl. Math. {\bf 28}, 391--401 (1989).

\bibitem{BoninEtAl2010}\vspace{-.3truecm}
C. A. Bonin, R. Bufalo, B. M. Pimentel and G. E. R. Zambrano,
\textit{Podolsky electromagnetism at finite temperature: Implications on the Stefan-Boltzmann law},
Phys. Rev.  {\bf D 81}, 025003 (2010).

\bibitem{AcciolyScatena2010}	\vspace{-.3truecm}
A. Accioly and E. Scatena,
\textit{Limits on the coupling constant of higher-derivative electromagnetism},
Mod. Phys. Lett.  {\bf A 25}, 269--276 (2010).

\bibitem{RSii}\vspace{-.3truecm}
M. Reed and B. Simon, 
{\it Fourier Analysis, Self-Adjoint\-ness} (Methods of Modern Mathematical Physics II),
(Acad. Press, Orlando, 1975).

\bibitem{RSiv}\vspace{-.3truecm}
M. Reed and B. Simon,
{\it Analysis of Operators}
(Methods of Modern Mathematical Physics IV),
(Acad. Press, Orlando, 1978).

\bibitem{AS}\vspace{-.3truecm}
M. Abramowitz and I. Stegun,
{\it Handbook of Mathematical Functions} 9th ed,
(Dover, New York, 1972).

\bibitem{Keppeler} \vspace{-.3truecm}
S. Keppeler,
\textit{Semiclassical quantisation rules for the Dirac and Pauli equations},
Annals Phys. (NY) {\bf 304}, 40--71 (2003).

\bibitem{PartheyEtAl2011}\vspace{-.3truecm}
C. G. Parthey et al.,
\textit{Improved measurement of the hydrogen 1S-2S transition frequency},
Phys. Rev. Lett. {\bf 107}, 203001 (2011).

\bibitem{LambRetherford1947}\vspace{-.3truecm}
W. E. Lamb and R. C. Retherford,
\textit{Fine structure of the hydrogen atom by a microwave method},
Phys. Rev. {\bf 72}, 241 (1947).

\bibitem{Bethe1947}\vspace{-.3truecm}
H. Bethe,
\textit{The electromagnetic shift of energy levels},
Phys. Rev. {\bf 72}, 339 (1947).

\bibitem{KiePerJMP}\vspace{-.3truecm}
M. K.-H. Kiessling and J. K. Percus,
\textit{Hard sphere fluids with chemical self-potentials,}
{J. Math. Phys.} {\bf 51}, 015206 (2010).

\bibitem{BrDrPRD}\vspace{-.3truecm}
S. J. Brodsky and S. D. Drell,
\textit{Anomalous magnetic moment and limits on fermion substructure},
Phys. Rev. {\bf D 22}, 2236--2243 (1980).

\bibitem{lorentzENCYCLOP}\vspace{-.3truecm}
H. A. Lorentz, 
\textit{Weiterbildung der Maxwellschen Theorie: Elektronentheorie},
{En\-cy\-kl\-op\"adie d. Mathematischen Wissenschaften} ${\bf V}2$,
Art. 14, 145--280 (1904).

\bibitem{Abraham}\vspace{-.3truecm}
M. Abraham, 
\textit{Prinzipien der Dynamik des Elektrons},
{Phys. Z.}{\bf 4}, 57--63,
{Annalen Phys.} {\bf 10}, 105--179 (1903).


\bibitem{BornA}\vspace{-.3truecm}
M. Born,
\textit{Modified field equations with a finite radius of the electron},
{Nature} {\bf 132}, 282 (1933).

\bibitem{BornInfeldBb}\vspace{-.3truecm}
M. Born and L. Infeld,
\textit{Foundation of the new field theory},
Proc. Roy. Soc. London {\bf A 144}, 425--451 (1934).

\bibitem{HellerMotz}\vspace{-.3truecm}
G. Heller and L. Motz,
\textit{Averages over portions of configuration space},
Phys. Rev. {\bf 46}, 502--505 (1934). 

\bibitem{GreinerETalA}\vspace{-.3truecm}
J. Rafelski, L. P. Fulcher and W. Greiner,
\textit{Superheavy elements and an upper limit to the electric field strength},
Phys. Rev. Lett. {\bf 27}, 958--961 (1971).

\bibitem{GreinerETalB}\vspace{-.3truecm}
G. Soff, J. Rafelski and W. Greiner,
\textit{Lower bound to limiting fields in nonlinear electrodynamics},
Phys. Rev. {\bf A7}, 903--907 (1973).

\bibitem{CarKiePRL}\vspace{-.3truecm}
H. Carley and M. K.-H. Kiessling,
\textit{Nonperturbative calculation of Born--Infeld effects on the Schr\"odinger spectrum of the hydrogen atom}, 
Phys. Rev. Lett. {\bf 96}, 030402 (1--4) (2006).

\bibitem{Franklin}\vspace{-.3truecm}
J. Franklin and T. Garon,
\textit{Approximate calculations of Born--Infeld effects on the relativistic hydrogen spectrum},
Phys. Lett. {\bf A 375}, 1391--1395 (2011).

\bibitem{CarKieMPAG}\vspace{-.3truecm}
H. Carley and M. K.-H. Kiessling,
\textit{Constructing graphs over $\Rset^n$ with small prescribed mean-curvature},
Math. Phys. Anal. Geom. {\bf 18}, 
25pp 
(DOI 10.1007/s11040-015-9177-6, 2015).

\bibitem{KieMBIinCMP} \vspace{-.3truecm}
M. K.-H. Kiessling,
\textit{On the quasi-linear elliptic PDE $-\nabla\cdot(\nabla{u}/\sqrt{1-|\nabla{u}|^2}) 
= 4\pi\sum_k a_k \delta_{s_k}$ in physics and geometry},
Commun. Math. Phys. {\bf 314}, 509--523 (2012); \textit{Correction:} 
Commun. Math. Phys. {\bf 364}, 825--833 (2018). 

\end{thebibliography}

\end{document}